\documentclass[aps,prd,onecolumn,groupedaddress,showpacs,nofootinbib,amssymb]{revtex4}
\usepackage[dvips]{graphicx}
\usepackage{amssymb}
\usepackage{amsmath}
\usepackage{graphicx}
\usepackage{amsfonts}
\usepackage{bm}
\usepackage{amsmath,amssymb,theorem}
\begin{document}

\title{Autonomous Dynamical System Approach for $f(R)$ Gravity}
\author{S.~D. Odintsov,$^{1,2}$\,\thanks{odintsov@ieec.uab.es}
V.K. Oikonomou,$^{3,4}$\,\thanks{v.k.oikonomou1979@gmail.com}}

\affiliation{$^{1)}$ ICREA, Passeig Luis Companys, 23, 08010 Barcelona, Spain\\
$^{2)}$ Institute of Space Sciences (IEEC-CSIC) C. Can Magrans s/n,
08193 Barcelona, Spain\\
$^{3)}$ Laboratory for Theoretical Cosmology, Tomsk State University
of Control Systems
and Radioelectronics, 634050 Tomsk, Russia (TUSUR)\\
$^{4)}$ Tomsk State Pedagogical University, 634061 Tomsk, Russia\\
}

\tolerance=5000

\begin{abstract}
In this work we shall investigate the cosmological dynamical system
of $f(R)$ gravity, by constructing it in such a way so that it is
rendered autonomous. We shall study the vacuum $f(R)$ gravity case,
but also the case that matter and radiation perfect fluids are
present along with the $f(R)$ gravity. The dynamical system is
constructed in such a way so that the time-dependence of the system
is contained in a single parameter which depends on the Hubble rate
and it's second derivative. The autonomous structure of the
dynamical system is achieved when this parameter is constant,
therefore we focus on these cases. For the vacuum $f(R)$ case, we
investigate two cases with the first leading to a stable de Sitter
attractor fixed point but also to an unstable de Sitter fixed point,
and the second is related to a matter dominated era. The stable de
Sitter attractor is also found for the $f(R)$ gravity in the
presence of matter and radiation perfect fluids. In all the cases we
performed a detailed numerical analysis of the dynamical system and
we also investigate in detail the stability of the resulting
fixed points. Also, we present an exceptional feature of the $R^2$
gravity model, in the absence of perfect fluids. Finally, we
investigate what is the approximate form of the $f(R)$ gravities
near the stable and the unstable de Sitter fixed points.
\end{abstract}

\pacs{95.35.+d, 98.80.-k, 98.80.Cq, 95.36.+x}

\maketitle

\section{Introduction}

Since the observation of the late-time acceleration
\cite{Riess:1998cb} that our Universe experiences, the focus in
modern theoretical cosmology is to find a theoretical framework that
can consistently harbor this late-time acceleration era. Many
theoretical proposals can potentially provide a consistent
description of the so-called dark energy era, with the modified
gravity being the main appealing approach, for reviews see
\cite{reviews1,reviews2,reviews3,reviews4}. Among the various
modified gravity descriptions, $f(R)$ gravity is the most important
and most studied approach, in the context of which it is possible to
provide the unification of the early with the late-time acceleration
of the Universe \cite{Nojiri:2003ft,Nojiri:2006gh}.

Due to the importance of $f(R)$ gravity, it is compelling to know
the general structure of the phase space of the cosmological
dynamical system of the theory. Many studies already exist in the
literature, studying various aspects of dynamical systems in
cosmology of modified gravity
\cite{Boehmer:2014vea,Bohmer:2010re,Goheer:2007wu,Leon:2014yua,Leon:2010pu,deSouza:2007zpn,Giacomini:2017yuk,Kofinas:2014aka,Leon:2012mt,Gonzalez:2006cj,Alho:2016gzi,Biswas:2015cva,Muller:2014qja,Mirza:2014nfa,Rippl:1995bg,Ivanov:2011vy,Khurshudyan:2016qox,Boko:2016mwr,Odintsov:2017icc},
see also \cite{Odintsov:2015wwp}, which focus on the existence of
fixed points and also their stability. In this paper we shall
attempt an alternative approach for the study of the $f(R)$ gravity
dynamical system case, in the presence or in the absence of perfect
matter fluids. Particularly, we shall investigate the phase space in
the limit that the corresponding dynamical system is strictly
autonomous. The motivation for studying the autonomous limit of the
cosmological dynamical system, comes mainly from the fact that if a
dynamical system is non-autonomous, then the stability of the fixed
points is not guaranteed by using the theorems which hold true in
the autonomous case. Hence, the stability of a fixed point
corresponding to a non-autonomous system is rather a problem without
a solution, unless highly non-trivial techniques are employed. Let
us explain our argument by exploiting a very simple example which
can be found in \cite{dynsystemsbook}, so consider the one
dimensional dynamical system $\dot{x}=-x+t$. The solution can be
easily found to be $x(t)=t-1+e^{-t}(x_0+1)$, from which it is
obvious that all the solutions asymptotically approach $t-1$ for
$t\to \infty$. Also it is easy to see that the only fixed point is
the time-dependent solution $x=t$, which however is not a solution
to the dynamical system. In addition, a standard analysis by using
the fixed point theorems, shows that the vector field actually move
away from the attractor $x(t)=t-1$, which is simply wrong.
Therefore, it is conceivable that knowing the fixed points of a
non-autonomous system, and then applying the Hartman-Grobman
theorems, does not provide a consistent solution to the stability
problem of the fixed points, and in effect, the structure of the
cosmological phase space cannot be revealed in such a way.

The purpose of this paper is to construct a dynamical system for
$f(R)$ gravity, and we aim to study the cases for which the
dynamical system is rendered autonomous. Particularly, we shall
quantify the whole time-dependence of the dynamical system in terms
of one single parameter $m$, which turns out to be equal to
$m=-\frac{\ddot{H}}{H^3}$, where $H$ is the Hubble rate of the
Universe. Then we will focus on various values of this parameter
with some physical significance, and specifically we focus on
$m\simeq 0$ and $m=-\frac{9}{2}$. The first case turns out to
describe an inflationary de Sitter final attractor which is stable,
in both the vacuum and matter fluid $f(R)$ gravity. We need to note
that not only the de Sitter or a quasi de Sitter cosmology yield
$m\simeq 0$, so the general study we perform covers all the
cosmologies which may yield $m\simeq 0$. This means that the case
$m\simeq 0$ may describe an approximate solution of some specific
$f(R)$ gravity, for example during the slow-roll era, or similar. In
all the cases, we shall find the fixed points of the dynamical
system, and the fact that the dynamical system is autonomous will
enable us to use the fixed point theorems and thus reveal the
stability structure of the system. Also we shall use various sets of
initial conditions, emphasizing for $e$-foldings numbers in the
range $N=[0,60]$, and we shall present in detail the stability
properties of the fixed points in terms of the resulting
trajectories. As a side study, we shall also consider in brief the
case $m=-\frac{9}{2}$, which as we show, describes a matter
dominated era, so we shall investigate the $f(R)$ gravity phase
space properties, and we also show that the stability of the matter
domination final attractor can be achieved. The latter study clearly
demonstrates that a  vacuum $f(R)$ gravity may successfully describe
a matter dominated era, however caution is needed for the stability
study of the fixed point. In principle there are other cases for
which the $f(R)$ gravity dynamical system can be rendered
autonomous, but we do not cover all the different cases for brevity.
We thoroughly discuss how an autonomous dynamical system can be
obtained in the context of $f(R)$ gravity  in the presence of matter
and radiation perfect fluids, and we mainly focus on de Sitter
attractors. Also, along with the stable de Sitter attractor, in the
vacuum $f(R)$ gravity case, there exists also an unstable de Sitter
fixed point. Intriguingly enough, as we explicitly demonstrate at a
later section, when a vacuum $R^2$ gravity is considered, the
dynamical system reaches the unstable de Sitter attractor. As we
argue, this feature may be a sign of graceful exit from inflation.
Also we shall investigate the approximate forms of the $f(R)$
gravities near the stable and the unstable de Sitter fixed points,
and also we discuss the role that play terms of the $R^2$ form, in
the graceful exit issue.

This paper is organized as follows: In section II, we present the
way to obtain an autonomous dynamical system from a general vacuum
$f(R)$ gravity. In section III, we study in detail the phase space
of vacuum $f(R)$ gravity, and we focus on the cases $m\simeq 0$ and
$m=-\frac{9}{2}$. We find the fixed points in each case and we
investigate the stability structure of the fixed points, by
analyzing many possible initial conditions. Moreover, we investigate
which are the approximate forms of the $f(R)$ gravities, near the
stable and the unstable de Sitter fixed points, and also we
thoroughly discuss the graceful exit issue, in view of the presence
of $R^2$ terms. Moreover, we present a specific example of an $f(R)$
cosmology, which satisfies $m\simeq 0$, and we examine the behavior
of the dynamical variables which were introduced in the previous
sections. In addition, other attractors apart from the de Sitter
one, are also discussed. In section IV we thoroughly study the
structure of the $f(R)$ gravity autonomous dynamical system, in the
presence of matter and radiation perfect fluids, emphasizing on the
de Sitter attractors. Finally, the conclusions follow in the end of
the paper.

Before starting our study, let us note that we will use a flat
Friedmann-Robertson-Walker (FRW) metric, as a geometric background,
the line element of which is,
\begin{equation}\label{frw}
ds^2 = - dt^2 + a(t)^2 \sum_{i=1,2,3} \left(dx^i\right)^2\, ,
\end{equation}
where $a(t)$ is the scale factor. Also for a FRW metric, the Ricci
scalar is,
\begin{equation}\label{ricciscalaranalytic}
R=6\left (\dot{H}+2H^2 \right )\, ,
\end{equation}
where $H=\frac{\dot{a}}{a}$, is the Hubble rate.

\section{The Autonomous Dynamical System of Vacuum $f(R)$ Gravity}

In this section we shall present the general structure of the vacuum
$f(R)$ gravity, and we will discuss the cases that may render this
an autonomous system. The vacuum $f(R)$ gravity action is,
\begin{equation}\label{action}
\mathcal{S}=\frac{1}{2\kappa^2}\int \mathrm{d}^4x\sqrt{-g}f(R)\, ,
\end{equation}
where $\kappa^2=8\pi G=\frac{1}{M_p^2}$ and also $M_p$ is the Planck
mass scale. In the context of the metric formalism, the equations of
motion are obtained by varying the gravitational action
(\ref{action}) with respect to the metric tensor $g_{\mu \nu}$, so
by  doing so we obtain the following equations of motion,
\begin{equation}\label{eqnmotion}
F(R)R_{\mu \nu}(g)-\frac{1}{2}f(R)g_{\mu
\nu}-\nabla_{\mu}\nabla_{\nu}f(R)+g_{\mu \nu}\square F(R)=0\, ,
\end{equation}
which can be written as follows,
\begin{align}\label{modifiedeinsteineqns}
R_{\mu \nu}-\frac{1}{2}Rg_{\mu
\nu}=\frac{\kappa^2}{F(R)}\Big{(}T_{\mu
\nu}+\frac{1}{\kappa^2}\Big{(}\frac{f(R)-RF(R)}{2}g_{\mu
\nu}+\nabla_{\mu}\nabla_{\nu}F(R)-g_{\mu \nu}\square
F(R)\Big{)}\Big{)}\, ,
\end{align}
with the prime indicating differentiation with respect to the Ricci
scalar. For the FRW metric of Eq. (\ref{frw}), the cosmological
equations of motion become,
\begin{align}
\label{JGRG15} 0 =& -\frac{f(R)}{2} + 3\left(H^2 + \dot H\right)
F(R) - 18 \left( 4H^2 \dot H + H \ddot H\right) F'(R)\, ,\\
\label{Cr4b} 0 =& \frac{f(R)}{2} - \left(\dot H + 3H^2\right)F(R) +
6 \left( 8H^2 \dot H + 4 {\dot H}^2 + 6 H \ddot H + \dddot H\right)
F'(R) + 36\left( 4H\dot H + \ddot H\right)^2 F'(R) \, ,
\end{align}
where $F(R)=\frac{\partial f}{\partial R}$, $F'(R)=\frac{\partial
F}{\partial R}$, and $F''(R)=\frac{\partial^2 F}{\partial R^2}$. In
order to reveal the autonomous structure of the cosmological
dynamical system described by the equations of motion
(\ref{JGRG15}), we shall introduce the following variables,
\begin{equation}\label{variablesslowdown}
x_1=-\frac{\dot{F}(R)}{F(R)H},\,\,\,x_2=-\frac{f(R)}{6F(R)H^2},\,\,\,x_3=
\frac{R}{6H^2}\, .
\end{equation}
In the following we shall use the $e$-foldings number $N$, instead
of the cosmic time, so the derivative with respect to the
$e$-foldings number can be expressed as follows,
\begin{equation}\label{specialderivative}
\frac{\mathrm{d}}{\mathrm{d}N}=\frac{1}{H}\frac{\mathrm{d}}{\mathrm{d}t}\,
,
\end{equation}
which shall be useful. Hence, by using the variables
(\ref{variablesslowdown}) and also the equations of motion
(\ref{JGRG15}), after some algebra we obtain the following dynamical
system,
\begin{align}\label{dynamicalsystemmain}
& \frac{\mathrm{d}x_1}{\mathrm{d}N}=-4-3x_1+2x_3-x_1x_3+x_1^2\, ,
\\ \notag &
\frac{\mathrm{d}x_2}{\mathrm{d}N}=8+m-4x_3+x_2x_1-2x_2x_3+4x_2 \, ,\\
\notag & \frac{\mathrm{d}x_3}{\mathrm{d}N}=-8-m+8x_3-2x_3^2 \, ,
\end{align}
where the parameter $m$ is equal to,
\begin{equation}\label{parameterm}
m=-\frac{\ddot{H}}{H^3}\, .
\end{equation}
By looking the dynamical system (\ref{dynamicalsystemmain}), it is
obvious that the only N-dependence (or time dependence) is contained
in the parameter $m$. Also we did not expressed $m$ as a function of
$N$, since we shall assume that this parameter will take constant
values. Hence if the parameter $m$ is a constant, the dynamical
system is rendered autonomous since no explicit time dependence
exist in it.

The effective equation of state (EoS) for a general $f(R)$ gravity
theory is,
\begin{equation}\label{weffoneeqn}
w_{eff}=-1-\frac{2\dot{H}}{3H^2}\, ,
\end{equation}
and it can be written in terms of the variable $x_3$ as follows,
\begin{equation}\label{eos1}
w_{eff}=-\frac{1}{3} (2 x_3-1)\, .
\end{equation}
By using the dynamical system (\ref{dynamicalsystemmain}) and the
EoS (\ref{eos1}), given the value of the parameter $m$, in the
following sections we shall investigate the structure of the phase
space corresponding to the vacuum $f(R)$ gravity, and we shall
discuss in detail the physical significance and implications of the
results.

\section{Study of the Vacuum $f(R)$ gravity Phase Space}

The parameter $m$ appearing in the non-linear dynamical system
(\ref{dynamicalsystemmain}) plays an important role, since it is the
only source of time-dependence in the dynamical system. Let us note
that for certain cosmological evolutions this parameter is constant.
For example, a quasi de Sitter evolution, in which case the scale
factor is,
\begin{equation}\label{quasidesitter}
a(t)=e^{H_0 t-H_i t^2}\, ,
\end{equation}
the parameter $m$ is equal to zero, and the same applies for a de
Sitter evolution. However, in this section we shall not assume that
the scale factor has a specific form, but we shall study in general
the cases $m\simeq 0$ and $m=-\frac{9}{2}$. As we shall see, these
two cases have special physical significance. With regard to the
$m\simeq 0$ case, this is easy to check, since if we solve the
differential equation $\frac{\ddot{H}}{H^3}=0$, this yields the
solution,
\begin{equation}\label{quasidesitterevolution1}
H(t)=H_0-H_i t\, ,
\end{equation}
This means that we focus on cosmologies for which the approximate
solution for the evolution is a quasi de Sitter evolution. This does
not mean that the exact Hubble rate is a quasi-de Sitter evolution,
but the approximate $f(R)$ gravity which drives the evolution, leads
to an approximate quasi-de Sitter evolution. Interestingly enough,
for the quasi-de Sitter evolution (\ref{quasidesitterevolution1}),
the following conditions hold true,
\begin{equation}\label{slowrollconditions}
H\dot{H}\gg \ddot{H},\,\,\,\dot{H}\ll H^2\, ,
\end{equation}
which are the slow-roll conditions. Hence the $m\simeq 0$ case is
related to the slow-roll condition on the inflationary era. In the
following sections we shall perform an in depth  analysis of the
$m\simeq 0$, since this is the most interest scenario.

Before discussing the details of the two cases  $m\simeq 0$ and
$m=-\frac{9}{2}$, let us present in brief the classical approach to
the classification of fixed points. According to the classical approach, the linearization
procedure applies \cite{dynsystemsbook}, which can also be applied in our case, if the fixed point is hyperbolic. Also we briefly discuss some essential features of the stability theory for dynamical systems .

With the stability theory of dynamical systems, basically one reveals the stability of the solutions and of the trajectories of dynamical systems. Basically, the stability of the solutions-trajectories, addresses the behavior of these, if small perturbations of the initial conditions are performed. Also it is important to have a solid grasp of the asymptotic behavior of the solutions and trajectories of a dynamical system, with asymptotic meaning after a long period of time. The duration and determination of this long period of time is usually determined by the physical scales in the theory. For the purposes of this paper, since we are interested in inflationary dynamics, the asymptotic behavior corresponds to approximately $N\simeq 60$ $e$-foldings. The simplest and most valuable behavior of the solutions-trajectories is exhibited by the fixed points of the dynamical system at hand, and also by periodic orbits. In all cases, how do orbits-trajectories or solutions behave near the fixed point, do these approach the fixed point, or alternatively said, are the trajectories attracted by the fixed point, or are these repelled from it? Also, if we perturb the orbit by using alternative initial conditions, how these perturbations affect the behavior of the trajectories near the fixed point? If the trajectory is always attracted by a fixed point, then this fixed point is called an attractor of the trajectories. Also if perturbed orbits tend asymptotically to a given orbit, then this orbit is called stable. The fixed points provide a characteristic picture of the structural stability of the dynamical system. The stability of a fixed point refers to the behavior of the trajectories near this point, if the trajectories are attracted to it, this is called stable (attractor), and if these are attracted to it asymptotically, this is called asymptotically stable (asymptotic attractor). In the case that trajectories are repelled from it, the fixed point is an unstable fixed point. One quantitative way to reveal the stability of a fixed point, is to apply the Hartman-Grobman theorem (see below), by linearizing the autonomous non-linear dynamical system. If the eigenvalues of the linearization matrix are negative real numbers, or even complex numbers with negative real parts, then the fixed point is stable (attractor). If none of the eigenvalues are purely imaginary, or equal to zero, then the fixed point can be an attractor or a repeller, and this is subject to the behavior of stable and unstable manifolds, see the book \cite{dynsystemsbook} for more details on this. The important thing to note is that the Hartman-Grobman theorem applies only on hyperbolic fixed points, which means that only in the case that the eigenvalues of the linearization matrix have non-zero real parts the Hartman-Grobman theorem may apply. It is the purpose of this paper to examine what happens in the cases that the Hartman-Grobman theorem does not apply.

As it is well known, the Hartman-Grobman linearization theorem
determines the stability and the structure of the phase space, when
hyperbolic fixed points are taken into account. Let $\Phi (t)$
$\epsilon$ $R^n$ be the solution to the following flow,
\begin{equation}\label{ds1}
\frac{\mathrm{d}\Phi}{\mathrm{d}t}=g(\Phi (t))\, ,
\end{equation}
with $g(\Phi (t))$  being a locally Lipschitz continuous map
$g:R^n\rightarrow R^n$. Let us denote with $\phi_*$ all the fixed
points of the dynamical system (\ref{ds1}), and the corresponding
Jacobian matrix, which we denote as $\mathcal{J}(g)$, is equal to,
\begin{equation}\label{jaconiab}
\mathcal{J}=\sum_i\sum_j\Big{[}\frac{\mathrm{\partial f_i}}{\partial
x_j}\Big{]}\, .
\end{equation}
The Jacobian has to be calculated at the fixed points, and the
eigenvalues $e_i$ must satisfy $\mathrm{Re}(e_i)\neq 0$. Let $\sigma
(A)$ denote the spectrum of the eigenvalues of $A$, so a hyperbolic
fixed point satisfies $\mathrm{Re}\left(\sigma
(\mathcal(J))\right)\neq 0$. The Hartman theorem ensures the
existence of a homeomorphism $\mathcal{F}:U\rightarrow R^n$, where
$U$ is an open neighborhood of $\phi_*$, such that
$\mathcal{F}(\phi_*)$. The homeomorphism generates a flow
$\frac{\mathrm{d}h(u)}{\mathrm{d}t}$, which is,
\begin{equation}\label{fklow}
\frac{\mathrm{d}h(u)}{\mathrm{d}t}=\mathcal{J}h(u)\, ,
\end{equation}
and in addition it is a topologically conjugate flow to the one
appearing in Eq. (\ref{ds1}). Sometimes the Hartman theorem holds
true for non-autonomous system, but the general rule is that it does
not apply to non-autonomous systems, or should be applied with
caution. A direct implication of the theorem is that the dynamical
system of Eq. (\ref{ds1}) may be written as follows,
\begin{equation}\label{dapprox}
\frac{\mathrm{d}\Phi}{\mathrm{d}t}=\mathcal{J}(g)(\Phi)\Big{|}_{\Phi=\phi_*}
(\Phi-\phi_*)+\mathcal{S}(\phi_*,t)\, ,
\end{equation}
where $\mathcal{S}(\phi,t)$ is a smooth map $[0,\infty )\times R^n$.
In effect, if the Jacobian $\mathcal{J}(g)$ satisfies
$\mathcal{Re}\left(\sigma (\mathcal{J}(g))\right)<0$ and also if
\begin{equation}\label{gfgd}
\lim_{\Phi\rightarrow
\phi_*}\frac{|\mathcal{S}(\phi,t)|}{|\Phi-\phi_*|}\rightarrow 0\, ,
\end{equation}
the fixed point $\phi_*$ corresponding to the dynamical flow
$\frac{\mathrm{d}\Phi}{\mathrm{d}t}=\mathcal{J}(g)(\Phi)\Big{|}_{\Phi=\phi_*
}(\Phi-\phi_*)$, is a fixed point of the flow in Eq.
(\ref{dapprox}), and it is stable asymptotically. It is conceivable
that if any of the conditions we presented above is violated,
further study of the phase space is needed in order to see whether
stability is ensure or not.

\subsection{de Sitter Inflationary Attractors and their Stability}

We start off with the case $m\simeq 0$, which may possibly describe
a quasi de Sitter evolution, however we shall analyze the dynamics
of the system (\ref{dynamicalsystemmain}), for $m\simeq 0$ without
specifying the Hubble rate.

In the case of the dynamical system of Eq.
(\ref{dynamicalsystemmain}), the matrix
$\mathcal{J}=\sum_i\sum_j\Big{[}\frac{\mathrm{\partial
f_i}}{\partial x_j}\Big{]}$ is equal to,
\begin{equation}\label{matrixceas}
\mathcal{J}=\left(
\begin{array}{ccc}
 2 x_1-x_3+3 & 0 & 2-x_1 \\
 x_2 & x_1-2 x_3+4 & -2 x_2-4 \\
 0 & 0 & 8-4 x_3 \\
\end{array}
\right)\, ,
\end{equation}
where the functions $f_i$ are,
\begin{align}\label{fis}
& f_1=-4-3x_1+2x_3-x_1x_3+x_1^2\, , \\ \notag &
f_2=8-4x_3+x_2x_1-2x_2x_3+4x_2 ,\\ \notag & f_3=8x_3-2x_3^2-8 \, .
\end{align}
The fixed points of the dynamical system
(\ref{dynamicalsystemmain}), for general $m$ are the following,
\begin{align}\label{fixedpointsgeneral}
& \phi_*^1=(\frac{1}{4} \left(-\sqrt{-2m}-\sqrt{-2 m+20 \sqrt{2}
\sqrt{-m}+4}-2\right),\frac{1}{4} \left(3 \sqrt{2}
\sqrt{-m}+\sqrt{-2 m+20 \sqrt{-2m}+4}-2\right),\frac{1}{2}
\left(4-\sqrt{-2m}\right)),
 \\ \notag & \phi_*^2=(\frac{1}{4} \left(-\sqrt{-2m}+\sqrt{-2 m+20 \sqrt{-2m}+4}-2\right),\frac{1}{4} \left(3 \sqrt{-2m}-\sqrt{-2 m+20 \sqrt{-2m}+4}-2\right),\frac{1}{2} \left(4-\sqrt{-2m}\right)),
 \\ \notag & \phi_*^3=(\frac{1}{4} \left(\sqrt{-2m}-\sqrt{-2 m-20 \sqrt{-2m}+4}-2\right),\frac{1}{4} \left(-3 \sqrt{-2m}+\sqrt{-2 m-20 \sqrt{-2m}+4}-2\right),\frac{1}{2} \left(\sqrt{-2m}+4\right))
 \\ \notag & \phi_*^4=(\frac{1}{4} \left(\sqrt{-2m}+\sqrt{-2 m-20 \sqrt{-2m}+4}-2\right),\frac{1}{4} \left(-\sqrt{2} \sqrt{-m-10 \sqrt{-2m}+2}-3 \sqrt{-2m}-2\right),\frac{1}{2} \left(\sqrt{2}
 \sqrt{-m}+4\right))\, .
\end{align}
In the case $m\simeq 0$, the fixed points are,
\begin{equation}\label{fixedpointdesitter}
\phi_*^1=(-1,0,2),\,\,\,\phi_*^2=(0,-1,2)\, .
\end{equation}
The eigenvalues for the fixed point $\phi_*^1$ are $(-1, -1, 0)$,
while for the fixed point $\phi_*^2$ these are $(1, 0, 0)$. Hence
both equilibria are non-hyperbolic, but as we show the fixed point
$\phi_*^1$ is stable and $\phi_*^2$ is unstable.

Before we proceed let us discuss the physical significance of the
two fixed points, and this can easily be revealed by observing that
in both the equilibria (\ref{fixedpointdesitter}), we have $x_3=2$.
By substituting $x_3=2$ in Eq. (\ref{eos1}), we get $w_{eff}=-1$, so
effectively we have two de Sitter equilibria.

Also it is worth to have a concrete idea on how the dynamical system behaves analytically. Actually, the third equation of the dynamical system (\ref{dynamicalsystemmain}) is decoupled, and the solution of it reads,
\begin{equation}\label{solutionanalyticx3}
x_3(N)=\frac{4 N-2 \omega +1}{2 N-\omega }\, ,
\end{equation}
where $\omega$ is an integration constant which can be fixed by the initial conditions. The asymptotic behavior of the solution (\ref{solutionanalyticx3}), that is for large $N$, is $x_3\to 2$, which is exactly the behavior we indicated earlier. Also, by using the solution (\ref{solutionanalyticx3}) and substituting it in the first two coupled differential equations of the dynamical system (\ref{dynamicalsystemmain}), we may obtain the analytic form of the solutions $x_1(N)$ and $x_2(N)$, which read,
\begin{align}\label{analyticformx1andx2}
& x_2(N)=c_1-\frac{\left(-c_1-2\right) c_2 U\left(1-\frac{-c_1-2}{c_1},2,N c_1-\frac{\omega  c_1}{2}\right)-c_1 L_{\frac{-c_1-2}{c_1}-1}^1\left(c_1 N-\frac{c_1 \omega }{2}\right)}{c_2 U\left(-\frac{-c_1-2}{c_1},1,N c_1-\frac{\omega  c_1}{2}\right)+L_{\frac{-c_1-2}{c_1}}\left(c_1 N-\frac{c_1 \omega }{2}\right)}, \\ \notag &
x_1(N)=-\frac{\left(-c_1-2\right) c_2 U\left(1-\frac{-c_1-2}{c_1},2,N c_1-\frac{\omega  c_1}{2}\right)-c_1 L_{\frac{-c_1-2}{c_1}-1}^1\left(c_1 N-\frac{c_1 \omega }{2}\right)}{c_2 U\left(-\frac{-c_1-2}{c_1},1,N c_1-\frac{\omega  c_1}{2}\right)+L_{\frac{-c_1-2}{c_1}}\left(c_1 N-\frac{c_1 \omega }{2}\right)}\, ,
\end{align}
where $c_1$ and $c_2$ are integration constants and also $U(a,b,z)$ and $L_n^a(x)$ are the confluent hypergeometric function and the generalized Laguerre polynomial respectively. In order to have a direct command on the initial conditions, we shall proceed directly on the numerical analysis, but it can be shown that if the integration constants are appropriately chosen, the numerical and analytical results may coincide.

\begin{figure}[h]
\centering
\includegraphics[width=20pc]{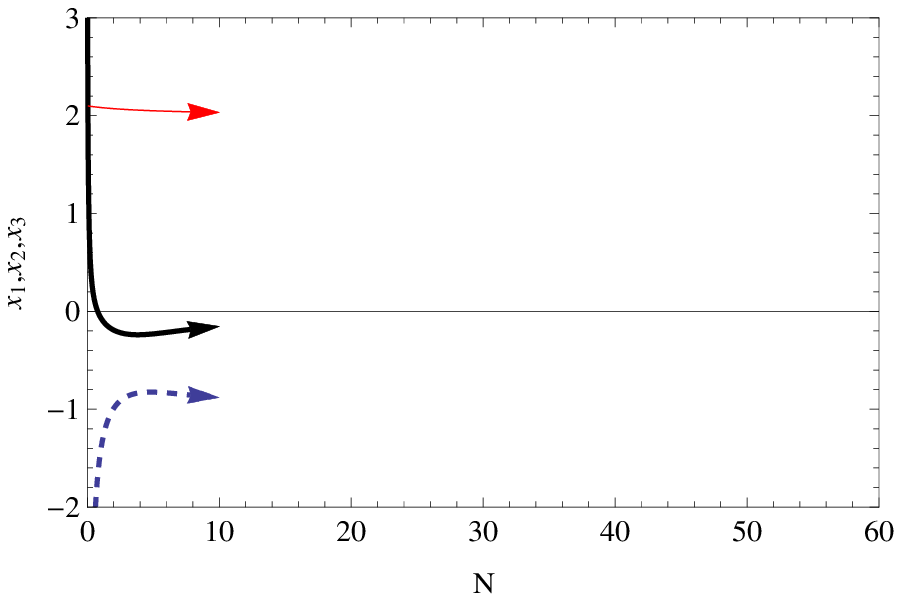}
\includegraphics[width=20pc]{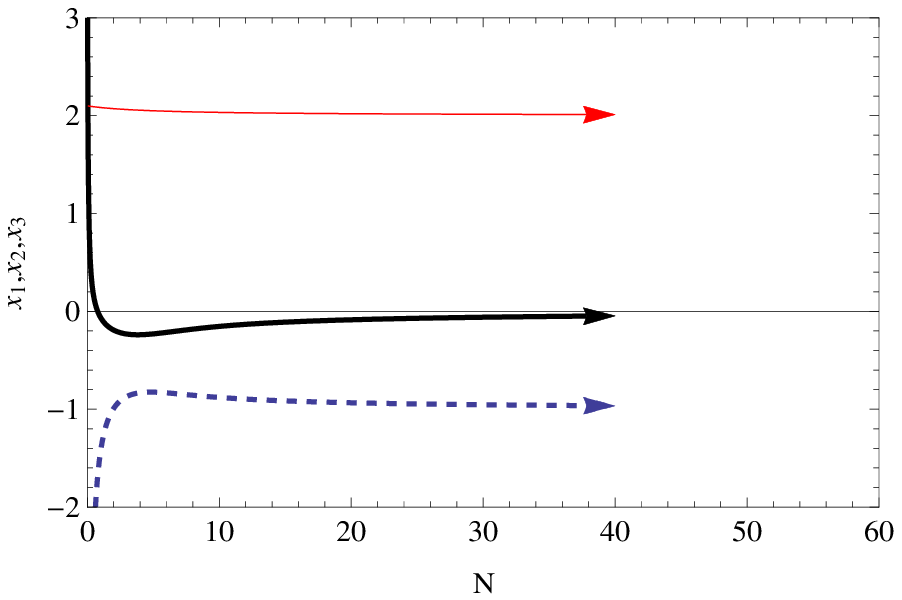}
\includegraphics[width=20pc]{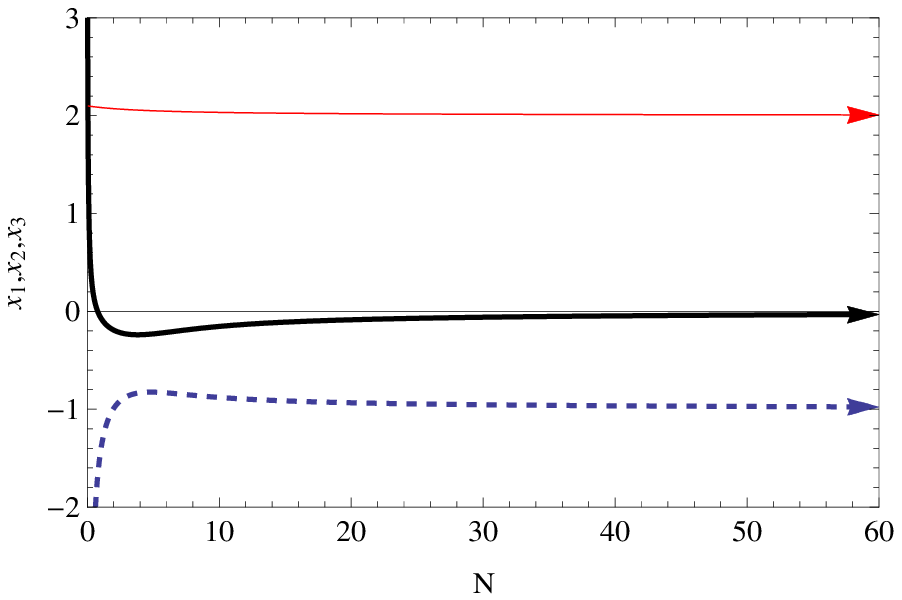}
\caption{{\it{Numerical solutions $x_1(N)$, $x_2(N)$ and $x_3(N)$
for the dynamical system (\ref{dynamicalsystemmain}), for the
initial conditions $x_1(0)=-8$, $x_2(0)=5$ and $x_3(0)=2.6$,  and
for $m\simeq 0$. }}} \label{plot1}
\end{figure}

\begin{figure}[h]
\centering
\includegraphics[width=20pc]{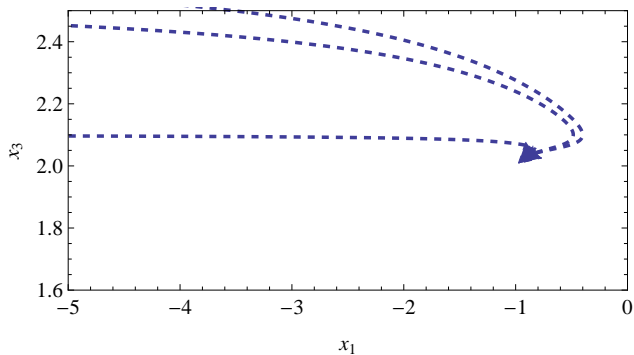}
\caption{{\it{Parametric plot in the plane $x_1-x_3$ for $N=(0,60)$,
for the dynamical system (\ref{dynamicalsystemmain}),  and for
$m\simeq 0$.}}} \label{plot2}
\end{figure}

\begin{figure}[h]
\centering
\includegraphics[width=20pc]{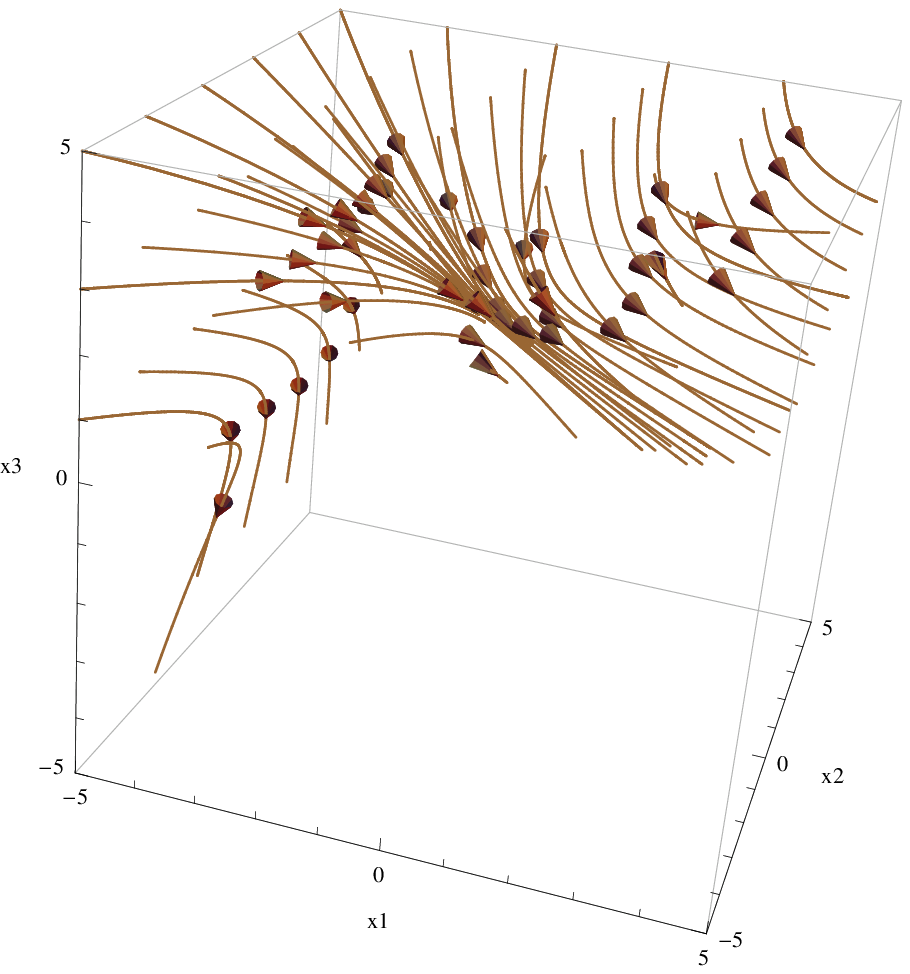}
\caption{{\it{Three dimensional flow for the dynamical system
(\ref{dynamicalsystemmain}), and for $m\simeq 0$.}}} \label{plot3}
\end{figure}

\begin{figure}[h]
\centering
\includegraphics[width=20pc]{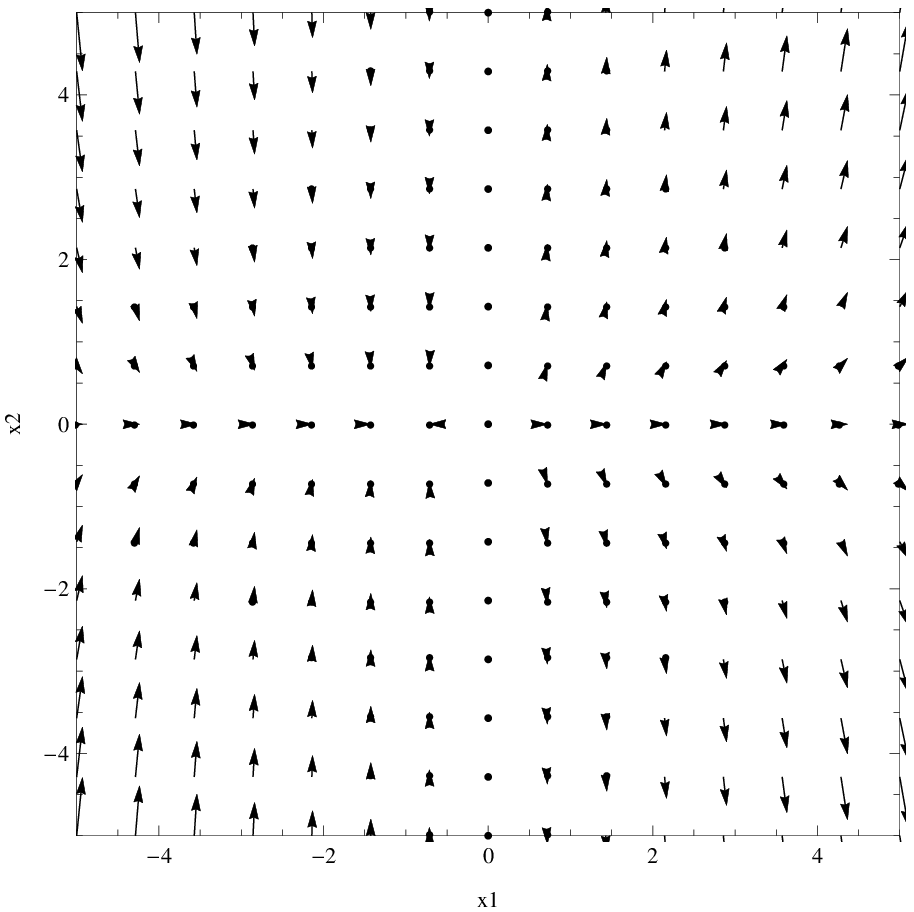}
\includegraphics[width=20pc]{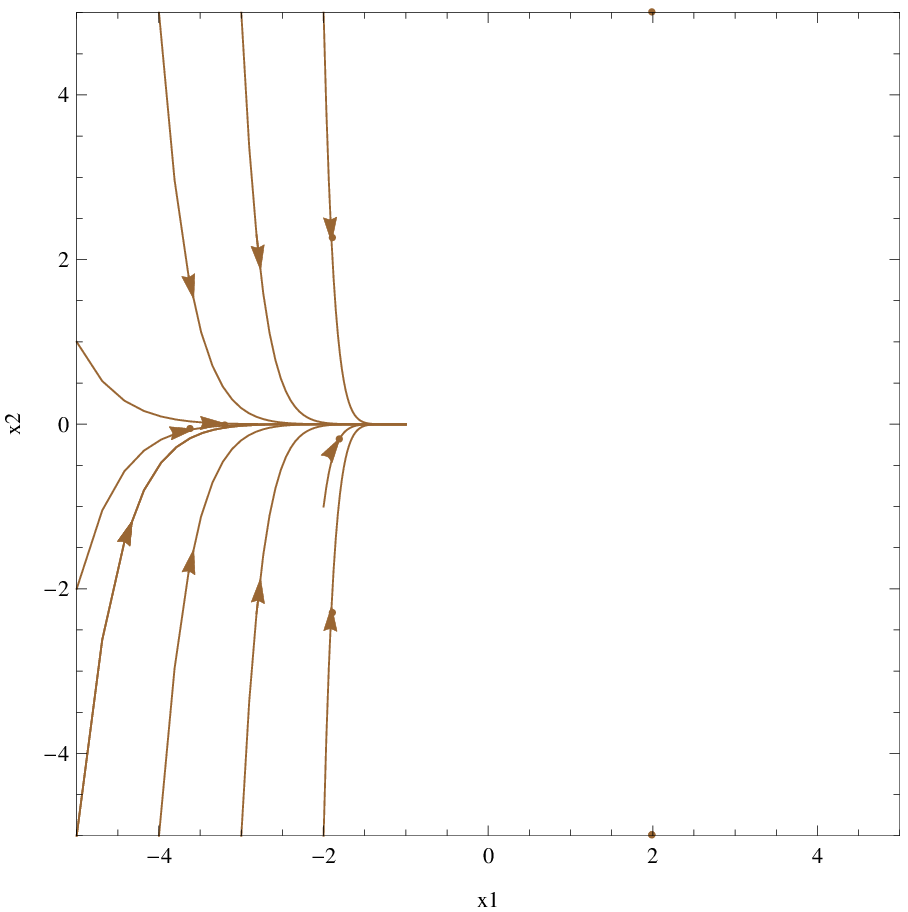}
\caption{{\it{The vector field flow (left) and the trajectories in
the $x_1-x_2$ plane (right)  for the dynamical system
(\ref{dynamicalsystemmain}), and for $m\simeq 0$.}}} \label{plot4}
\end{figure}

Now let us analyze the dynamics of the cosmological system, and for starters we numerically
solve the dynamical system (\ref{dynamicalsystemmain}) for various
initial conditions and with the $e$-foldings number belonging to the
interval $N=(0,60)$. In Fig. (\ref{plot1}) we present the numerical
solutions for the dynamical system (\ref{dynamicalsystemmain}), for
the initial conditions $x_1(0)=-8$, $x_2(0)=5$ and $x_3(0)=2.6$.
For the upper left plot of Fig. \ref{plot1}, the $e$-foldings number
is chosen in the interval $N=(0,10)$ and as it can be seen, the
equilibrium $\phi_*^1=(x_1,x_2,x_3)=(-1,0,2)$ is not reached yet,
and the same applies for the upper right plot in Fig. \ref{plot1},
in which case $N=(0,40)$, although the situation is a bit better. In
the third plot of Fig. \ref{plot1}, the $e$-foldings number is
chosen in the interval $N=(0,60)$, so by $N\sim 60$, the equilibrium
$\phi_*^1$ is reached. The same can be seen in Fig. (\ref{plot2}),
where we plot $x_1-x_3$, for various initial conditions.
From Figs. \ref{plot1} and \ref{plot2} it already seems that the
fixed point $\phi_*^1$ is a stable de Sitter attractor. As we show
by using further numerical analysis, indeed this  de Sitter
equilibrium is the final attractor of the phase space. In Fig.
\ref{plot3}, we plot the three dimensional flow, and as it can be
seen, for various initial conditions, the global attractor behavior
of $\phi_*^1$ can be seen. Also the stability of the attractor
$\phi_*^1$ can be seen in the reduced dynamical system for $x_3=2$.
Actually, the equation for $x_3$ can be solved directly since this
is decoupled from the rest of the equations in
(\ref{dynamicalsystemmain}).
 The dynamics of the reduced system for $x_3=2$ are
presented in Fig. \ref{plot4}, where we plot the vector field flow
(left) and the trajectories in the $x_1-x_2$ plane (right), for
various initial conditions. It is obvious that the attractor
behavior of the de Sitter fixed point $\phi_*^1$ is further
supported by these flows in the reduced system.
In conclusion, the case $m\simeq 0$ for the dynamical system
(\ref{dynamicalsystemmain}) results to a stable de Sitter attractor
and also to an unstable de Sitter fixed point. This result is
particularly interesting since it dictates that a general $f(R)$
gravity theory with Hubble rate that satisfies $m\simeq 0$ results
to an asymptotic stable de Sitter attractor, which is reached at
$N\sim 50-60$, as it can be seen in Fig. \ref{plot1}. This is quite
appealing, if one considers an inflationary interpretation of the
theory, for example if the scale factor is assumed to be a quasi-de
Sitter one, namely $a(t)\sim e^{H_0 t-H_i t^2}$. In the quasi-de
Sitter case, the attractor property reveals that the phase space of
a the $f(R)$ gravity has an asymptotic de Sitter attractor, which is
stable, which clearly may be interpreted having an inflationary
evolution in mind. However, not only a quasi-de Sitter cosmology
results to $m\simeq 0$, consider for example the symmetric bounce
cosmology \cite{Odintsov:2015ynk}, in which case the scale factor is
$a(t)=e^{\lambda t^2}$, with $\lambda>0$. In the symmetric bounce,
the parameter $m$ is also zero, but in this case the physical
interpretation of the de Sitter attractor is different, since it can
be a late-time de Sitter attractor. However, this requires different
initial conditions and perhaps a re-parametrization of the dynamical
system (\ref{dynamicalsystemmain}), in terms of the cosmic time, in
order to avoid inconsistencies. We leave this task as a possible
future work. Also, it is possible that for certain cosmological
models, we may have asymptotically $m\simeq 0$, if the parameters of
the theory are appropriately chosen. In such case, the
interpretation of the fixed points needs special attention.

Let us note that there is also an unstable de Sitter fixed point,
namely $\phi_*^2=(0,-1,2)$. So it may be possible that there are
$f(R)$ gravities which may lead the dynamical system directly on the
unstable de Sitter attractor. This can be seen in the 3-dimensional
flow of Fig. \ref{plot3}, since there are trajectories that move
away from the stable de Sitter attractor $\phi_*^1$. One example of
this sort, we shall present at a later section.

Now the central question is, what is the physical interpretation of
the above dynamics, and to which $f(R)$ gravities the case $m\simeq
0$ corresponds. It is conceivable that we can find approximate
expressions of the $f(R)$ gravities near the fixed points,
$\phi_*^1=(-1,0,2)$ and $\phi_*^2=(0,-1,2)$. This is the subject of
the next section.

\subsubsection{Approximate Form of the $f(R)$ Gravities Near the de Sitter Attractors}

In this section we shall analyze the physical implications of the
dynamics we presented in the previous section. As we demonstrate,
for the value of $m$ for which $m\simeq 0$, there are two de Sitter
fixed points, one stable, which is $\phi_*^1=(-1,0,2)$ and also an
unstable, which is $\phi_*^2=(0,-1,2)$. Now we shall investigate the
behavior of the $f(R)$ gravity near the fixed points and for
$m\simeq 0$. Recall that as we discussed in Eq.
(\ref{slowrollconditions}), the case $m\simeq 0$ is equivalent to
the slow-roll approximation, so effectively what we seek for is the
behavior of the $f(R)$ gravities near the fixed points and with the
slow-roll approximation holding true. Let us start with the first
fixed point, namely  $\phi_*^1=(-1,0,2)$, so the following
differential equations must hold true simultaneously at the fixed
point,
\begin{align}\label{caseidiffseqns}
-\frac{\mathrm{d}^2f}{\mathrm{d}R^2}\frac{\dot{R}}{H\frac{\mathrm{d}f}{\mathrm{d}R}}\simeq
-1,\,\,\,\frac{f}{H^2\frac{\mathrm{d}f}{\mathrm{d}R}6}\simeq 0\, ,
\end{align}
which stem from the conditions $x_1\simeq -1$ and $x_2\simeq 0$.
Since $m\simeq 0$ (or equivalently since the slow-roll approximation
holds true), the left differential equation can be written as
follows,
\begin{equation}\label{diifffrexp}
-24
H_i\frac{\mathrm{d}^2f}{\mathrm{d}R^2}-\frac{\mathrm{d}f}{\mathrm{d}R}=0\,
,
\end{equation}
which can easily be solved and it yields,
\begin{equation}\label{solapprox1}
f(R)\simeq \Lambda_1-24 \Lambda_2e^{-\frac{R}{24H_i}}\, .
\end{equation}
The $f(R)$ gravity solution (\ref{solapprox1}) is nothing but the
approximate form of the $f(R)$ gravity in the large curvature era,
which generates the quasi-de Sitter evolution of Eq.
(\ref{quasidesitterevolution1}) or equivalently, that yields
$m\simeq 0$. So since we are interested in the large curvature era,
the exponential term is subdominant, so the cosmological constant
$\Lambda_1$ leads to the stable de Sitter attractor
$\phi_*^1=(-1,0,2)$. By using the solution (\ref{solapprox1}), it is
easy to show that the second condition in Eq. (\ref{caseidiffseqns})
can be written as follows,
\begin{equation}\label{asxet1}
\frac{H_i \Lambda_1 e^{\frac{H_0^2}{2 H_i}}}{6 H_0^2
\Lambda_2}\simeq 0\, ,
\end{equation}
so it follows that in order the above condition is true, we must
have $H_0^2\gg 1$ and also $H_0^2\Lambda_2 \gg H_i \Lambda_1
e^{\frac{H_0^2}{2 H_i}}$. Notice that the final form of the $f(R)$
gravity is very similar to the exponential $f(R)$ gravity models of
Ref. \cite{Elizalde:2010ts}, in which case the full $f(R)$ gravity
is,
\begin{equation}\label{newadd1}
f(R)=R-2 \Lambda (1-e^{-2\frac{R}{R_0}})\, .
\end{equation}

Now let us consider the case of the second de Sitter fixed point,
namely $\phi_*^2=(0,-1,2)$, and in this case the conditions
$x_1\simeq 0$ and  $x_2\simeq -1$ become,
\begin{align}\label{caseidiffseqns1}
-\frac{\mathrm{d}^2f}{\mathrm{d}R^2}\frac{\dot{R}}{H\frac{\mathrm{d}f}{\mathrm{d}R}}\simeq
0,\,\,\,-\frac{f}{H^2\frac{\mathrm{d}f}{\mathrm{d}R}6}\simeq -1\, .
\end{align}
By using the fact that  $R\simeq 12 H^2$, when the quasi-de Sitter
evolution is taken into account, the second differential equation
can be written,
\begin{equation}\label{seconddiff}
f\simeq \frac{\mathrm{d}f}{\mathrm{d}R} \frac{R}{2}\, ,
\end{equation}
which can be solved to yield,
\begin{equation}\label{approximatersquare}
f(R)\simeq \alpha R^2\, .
\end{equation}
The solution (\ref{approximatersquare}) is not the exact form of the
$f(R)$ gravity which leads the cosmological system to the fixed
point, but it is the approximate form of the $f(R)$ gravity near the
fixed point $\phi_*^1$ which corresponds to the case $m\simeq 0$.
The approximate $f(R)$ gravity of Eq. (\ref{approximatersquare}) is
very similar to the $R^2$ model. Finally, the first condition in Eq.
(\ref{caseidiffseqns1}), for the $f(R)$ gravity being of the form
(\ref{approximatersquare}), becomes,
\begin{equation}\label{condition2extrasecond}
\frac{H_i}{H_0^2}\simeq 0\, ,
\end{equation}
which holds true if $H_0^2\gg H_i$.

Thus by taking into account the resulting forms of the $f(R)$
gravities we discussed in this section, we may conclude that the
$R^2$ terms related to quasi-de Sitter solutions always lead to an
unstable de Sitter fixed attractor, while terms containing
cosmological constants and exponentials, lead to a stable de Sitter
attractor. This result is interesting, since it is well known
\cite{Bamba:2014jia} that $R^2$ corrections to viable $f(R)$
gravities, like the exponential, always trigger graceful exit from
inflation, see the well-known viable Starobinsky inflation model \cite{starobinsky}. Hence, the presence of an $R^2$ term
leads to instability, which may be viewed as an indication of the
graceful exit from the inflationary era. In the following section we
shall analyze in detail the case of the $R^2$ gravity, and we will
investigate whether the predicted behavior we found in the previous
sections, holds true in this case.

\subsubsection{Specific Example I: $R^2$ Gravity and Graceful Exit from Inflation}

\begin{figure}[h]
\centering
\includegraphics[width=20pc]{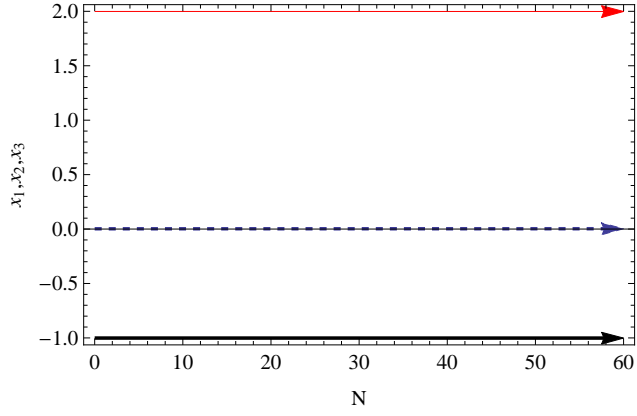}
\caption{{\it{The behavior of $x_1(N)$, $x_2(N)$ and $x_3(N)$ for
$H_0=10^{13}$sec$^{-1}$ and $H_i=10^{20}$sec$^{-2}$. The red curve
corresponds to $x_3(N)$, the black to $x_2(N)$ and finally the
dashed blue curve corresponds to the variable $x_1(N)$}}}
\label{plotextra}
\end{figure}

An intriguing example that leads to the picture we discussed in the
previous section, occurs for the $R^2$ gravity, with regards to the
final unstable de Sitter fixed point. As we will show, this $f(R)$
gravity makes the dynamical variables to end up to the unstable de
Sitter fixed point. The functional form of the $R^2$ gravity is,
\begin{equation}\label{r2gtravity}
f(R)=R+\frac{R^2}{36H_i}\, ,
\end{equation}
where the parameter $H_i$ has dimensions of mass$^2$. We shall focus
on the vacuum $f(R)$ gravity case in the rest of this section. In
order to investigate the behavior of the dynamical variables
(\ref{variablesslowdown}), we shall need the exact functional form
of the Hubble rate, and also this technique will hold true only if
the specific Hubble rate yields $m\simeq 0$. This issue needs to be
further explained at this point. The behavior of the dynamical
variables we described in the previous sections, holds true for all
the cosmologies which yield $m\simeq 0$, without specifying the
exact form of the $f(R)$ gravity, however the result shows us that
there are $f(R)$ gravities, actually those of Eq.
(\ref{solapprox1}), which yield a Hubble rate for which $m\simeq 0$
and also the variables $x_1$, $x_2$ and $x_3$ approach the values
$(x_1,x_2,x_3)=(-1,0,2)$, which correspond to the fixed point
$\phi_*^1$. Also there are $f(R)$ gravities which lead to an
unstable de Sitter fixed point, which can be approximated near the
fixed point as in Eq. (\ref{approximatersquare}). However, finding
the Hubble rate for a general $f(R)$ gravity functional form, is
quite a formidable task. Nevertheless, for the $R^2$ gravity
(\ref{r2gtravity}) this is easy when the slow-roll approximation is
used. The cosmological equations for the $R^2$ gravity in a FRW
background metric, are,
\begin{equation}
\label{frweqnsr2} \ddot{H}-\frac{\dot{H}^2}{2H}+3H_iH=-3H\dot{H}\,
,\quad \ddot{R}+3HR+6H_iR=0\, ,
\end{equation}
so in the slow-roll case, since $\ddot{H}\ll H\dot{H} $ and
$\dot{H}\ll H^2$, the second equation of Eq. (\ref{frweqnsr2}) can
be solved to yield the following Hubble rate,
\begin{equation}
\label{quasievolbnexampler2} H(t)\simeq H_0-H_i(t-t_k)\, ,
\end{equation}
which is a quasi-de Sitter evolution. As it can be easily checked,
and we also discussed this issue in the previous sections, the
Hubble rate (\ref{quasievolbnexampler2}) yields $m=0$. By using the
Hubble rate (\ref{quasievolbnexampler2}), the functional form of the
$R^2$ gravity (\ref{r2gtravity}) and also by expressing the cosmic
time as a function of the $e$-foldings number $N$ (recall that
$N=\ln a$), the dynamical variables (\ref{variablesslowdown}) as a
function of the $e$-foldings number $N$ become,
\begin{align}\label{efoldignsdynamicalvariables111}
& x_1(N)=\frac{2 H_i}{H_0^2-2 H_i N+H_i}, \\
\notag & x_2(N)=-\frac{\left(2 H_0^2+H_i (5-4 N)\right) \left(2
H_0^2-H_i (4 N+1)\right)}{4 \left(H_0^2-2 H_i N\right) \left(H_0^2-2
H_i N+H_i\right)},\\ \notag & x_3(N)=\frac{2 H_0^2-4 H_i
N-H_i}{H_0^2-2 H_i N}\, .
\end{align}
As it was shown in \cite{Odintsov:2015gba}, for the choice
$H_0=10^{13}$sec$^{-1}$ and $H_i=10^{20}$sec$^{-2}$, a viable
inflationary cosmology can be generated, so by using the
aforementioned values, in Fig. \ref{plotextra} we plot the behavior
of the variables (\ref{efoldignsdynamicalvariables111}) as a
function of the $e$-foldings number. The red curve corresponds to
$x_3(N)$, the black to $x_2(N)$ and finally the blue dashed curve
corresponds to the variable $x_1(N)$.
As it can be seen, the variables $x_1(N)$, $x_2(N)$ and $x_3(N)$
approach quite fast the fixed point values $(x_1,x_2,x_3)=(0,-1,2)$,
and therefore the $R^2$ gravity is an example of an $f(R)$ gravity
which leads to an unstable de Sitter fixed point. Actually this is
quite intriguing, since the unstable de Sitter fixed point is
reached for the $R^2$ gravity, and not the stable de Sitter point.
This feature could be seen as an indication that the graceful exit
from inflation occurs for the $R^2$ gravity. In principle, the
graceful exit issue cannot be addressed by studying solely the
variables $x_i$, $i=1,2,3$, since this may occur due to curvature
perturbations, so it is an effect which can be seen in higher orders
of curvature. For a detailed analysis on this, see
\cite{Odintsov:2015wwp}. However, the $R^2$ gravity shows an
exceptional behavior, since for this case the unstable de Sitter
fixed point is reached, and this is not accidental, since as we
discussed earlier, $R^2$ terms are always related to the graceful
exit from inflation, see \cite{Bamba:2014jia,starobinsky}.

Let us briefly discuss our claim that connects the graceful exit and the $R^2$ gravity. In order to be more concise, there is indication for the graceful exit from inflation in the $R^2$ gravity case, thus our claim is more mild in rigidity. According to Refs. \cite{Bamba:2014jia,Odintsov:2015wwp}, where a deeper analysis was performed, the graceful exit from inflation may be triggered by growing curvature perturbations around an unstable de Sitter solution of the theory. In this case, the de Sitter solution ceases to be the final attractor of the theory, and thus inflation ends as a consequence of that, since the de Sitter solution is not the final attractor of the theory. What follows is part of the subsequent physical evolution, for example the reheating era, since the slow-roll era ends. In order to investigate the perturbations of the curvature, we perturb the de Sitter solution $H=H_0$ as follows,
\begin{equation}\label{perturbationdesitter}
H(t)=H_0+\Delta H\, ,
\end{equation}
and we plug the perturbation in the first equation of motion of Eq. (\ref{frweqnsr2}), and by keeping linear terms of the derivatives of $\Delta H$, we obtain the following differential equation,
\begin{equation}\label{perturbationequation1}
3 H_0 H_i+H_0 \Delta \ddot{H}(t)+3 H_0 \Delta \dot{H}(t)+3 H_i \Delta H(t)+\frac{\Delta \dot{H}(t)}{2}=0
\end{equation}
which can be solved analytically and the solution is,
\begin{align}\label{solutionperturbations}
& \Delta H(t)=c_1 \exp \left(\frac{t \left(-\sqrt{36 H_0^2-48 H_0 H_i+12 H_0+1}-6 H_0-1\right)}{4 H_0}\right)\\ \notag &+c_2 \exp \left(\frac{t \left(\sqrt{36 H_0^2-48 H_0 H_i+12 H_0+1}-6 H_0-1\right)}{4 H_0}\right)-H_0\, ,
\end{align}
where $c_1$ and $c_2$ are integration constants. As it can be seen from the solution of Eq. (\ref{solutionperturbations}), the term proportional to $c_2$ is growing exponentially as a function of the cosmic time, therefore this supports our claim that the de Sitter point of the $R^2$ gravity is unstable, which we also demonstrated earlier by using the fixed point analysis. Basically, this unstable de Sitter point is the fixed point $(x_1,x_2,x_3)=(0,-1,2)$. We need to stress once more that this is an indication that graceful exit is triggered by curvature perturbations, however this indication cannot be considered as a proof of the graceful exit from inflation.

\subsection{Other Possible Attractors and their Stability}

Let us now briefly discuss the case $m=-\frac{9}{2}$, which as we
show corresponds to a matter dominated Universe. In the case at
hand, the matrix
$\mathcal{J}=\sum_i\sum_j\Big{[}\frac{\mathrm{\partial
f_i}}{\partial x_j}\Big{]}$ is equal to,
\begin{equation}\label{matrixceas1}
\mathcal{J}=\left(
\begin{array}{ccc}
 2 x_1-x_3+3 & 0 & 2-x_1 \\
 x_2 & x_1-2 x_3+4 & -2 x_2-4 \\
 0 & 0 & 8-4 x_3 \\
\end{array}
\right)\, ,
\end{equation}
where the functions $f_i$ are in this case,
\begin{align}\label{fis1}
& f_1=x_1^2-x_1 x_3+3 x_1+2 x_3-4\, ,
\\ \notag & f_2=-\frac{9}{2}+x_1 x_2-2 x_2 x_3+4 x_2-4 x_3+8,\\ \notag &
f_3=\frac{9}{2}-2 x_3^2+8 x_3-8 \, .
\end{align}
The fixed points in this case are,
\begin{equation}\label{fixedpointdesitter1}
\phi_*^1=(\frac{1}{4} \left(-5-\sqrt{73}\right),\frac{1}{4}
\left(7+\sqrt{73}\right),\frac{1}{2}),\,\,\,\phi_*^2=(\frac{1}{4}
\left(\sqrt{73}-5\right),\frac{1}{4}
\left(7-\sqrt{73}\right),\frac{1}{2})\, .
\end{equation}
For the first fixed point $\phi_*^1$ the eigenvalues are
$(6,-4.272,-0.386001)$, while for the second fixed point $\phi_*^2$
these are $(6,4.272,3.886)$. Both equilibria are non-hyperbolic and
the first seems to be stable. However as we show, if the initial
conditions on the variable $x_3$ deviate from the equilibrium value
$x_3=\frac{1}{2}$, the dynamical system solutions simply blow-up.

The fixed points have some physical interpretation, since for
$x_3=\frac{1}{2}$, the EoS in Eq. (\ref{eos1}) becomes  $w_{eff}=0$,
which perfectly describes a matter dominated era. What now remains
is to investigate the dynamics of the cosmological system.

As in the de Sitter case, it is possible to solve analytically the dynamical system, and in this case the solution for $x_3$ is,
\begin{align}\label{analyticformx1andx211}
& x_3(N)=\frac{7 e^{6 N}-\omega }{2 \left(e^{6 N}-\omega \right)},
\end{align}
where $\omega$ is again an integration constant. The solutions for $x_2$ and $x_3$ can also be found analytically, but the general solutions are particularly complicated and lengthy to include them at this point. From the solution (\ref{analyticformx1andx211}) it is obvious that asymptotically, $x_3\simeq 7/2$ which does not correspond to the matter dominated era value of $x_3$. Actually, this corresponds to an additional fixed point, which we shall consider below, however it has no physical significance, since it corresponds to a phantom evolution.

\begin{figure}[h]
\centering
\includegraphics[width=20pc]{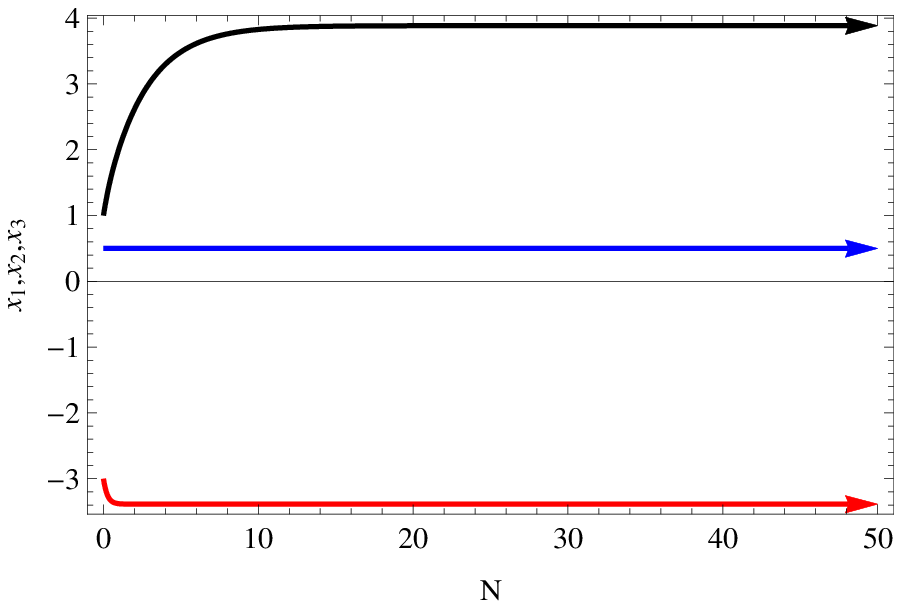}
\includegraphics[width=20pc]{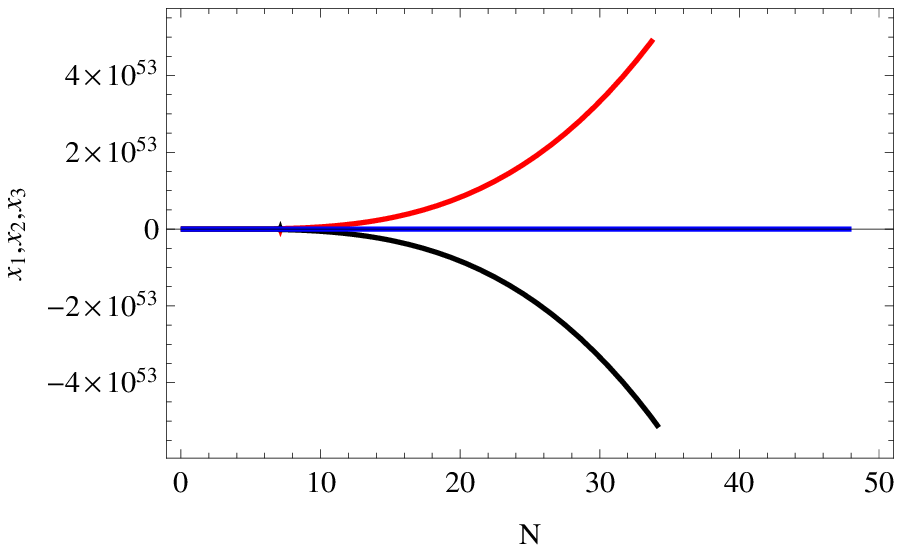}
\caption{{\it{The behavior of the solutions for the initial
conditions $x_1(0)=-3$, $x_2(0)=1$ and $x_3(0)=0.5$ (left plot) and
for $x_1(0)=-3$, $x_2(0)=1$ and $x_3(0)=0.501$ (right plot), for
$m=-9/2$.}}} \label{plot5}
\end{figure}

Now we proceed to the dynamical system analysis, and as we now show, if the initial conditions on $x_3$ deviate from the
equilibrium value, the solutions tend to infinity. In Fig.
\ref{plot5}, we plotted the behavior of the solutions for the
initial conditions $x_1(0)=-3$, $x_2(0)=1$ and $x_3(0)=0.5$ (left
plot) and for $x_1(0)=-3$, $x_2(0)=1$ and $x_3(0)=0.501$ (right
plot).
As it can be seen in Fig. \ref{plot5}, if the variable $x_3$
deviates from it's equilibrium value, the solutions blow up and the
fixed point is not a stable attractor. Hence, the stability of the
system is ensured only if $x_3=\frac{1}{2}$, and only for this case
$w=0$. Now we need to interpret this physically, and the
interpretation is simple, since the behavior of the dynamical system
indicates that when $x_3$ deviates from it's equilibrium value
$x_3=\frac{1}{2}$, which is equivalent to say when $w_{eff}\neq 0$,
then the vacuum $f(R)$ gravity does not approach the first fixed
point $\phi_*^1$. Actually, this is reasonable, since it seems that
the vacuum $f(R)$ gravity fails to describe a matter dominated era
unless $w_{eff}=0$. To our opinion, the matter dominated attractor
is reached only if $x_3=\frac{1}{2}$, and if this condition is not
met, instability occurs and the system ceases to describe a matter
dominated era.

\begin{figure}[h]
\centering
\includegraphics[width=20pc]{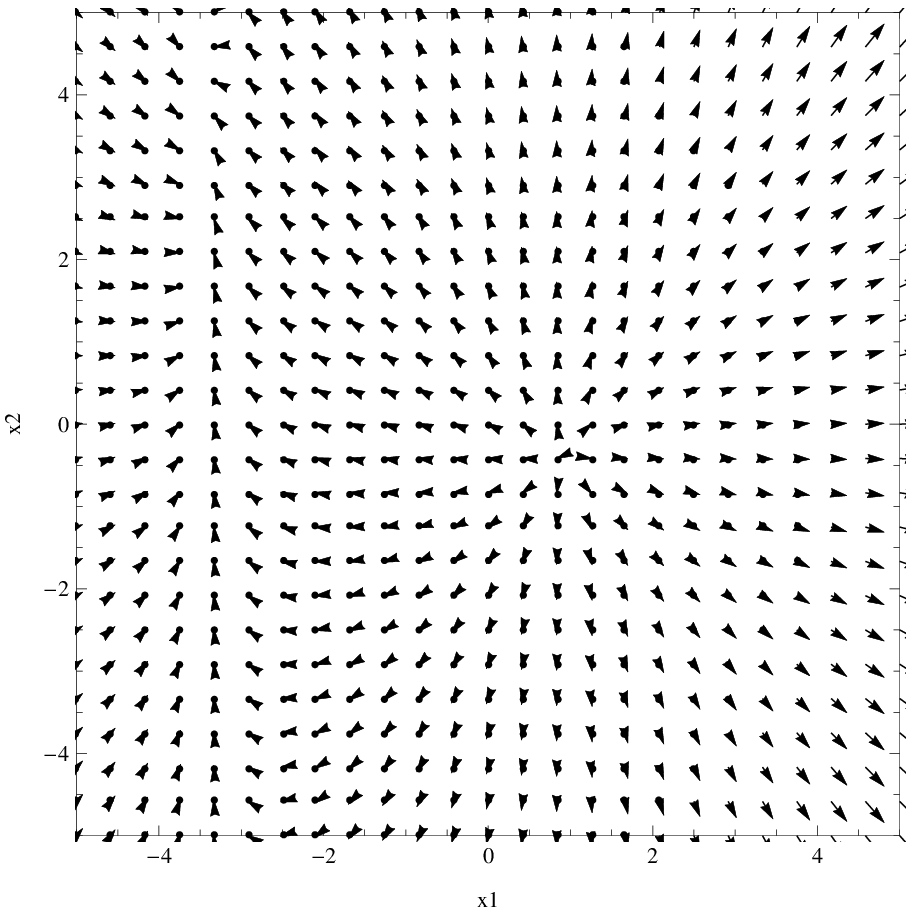}
\includegraphics[width=20pc]{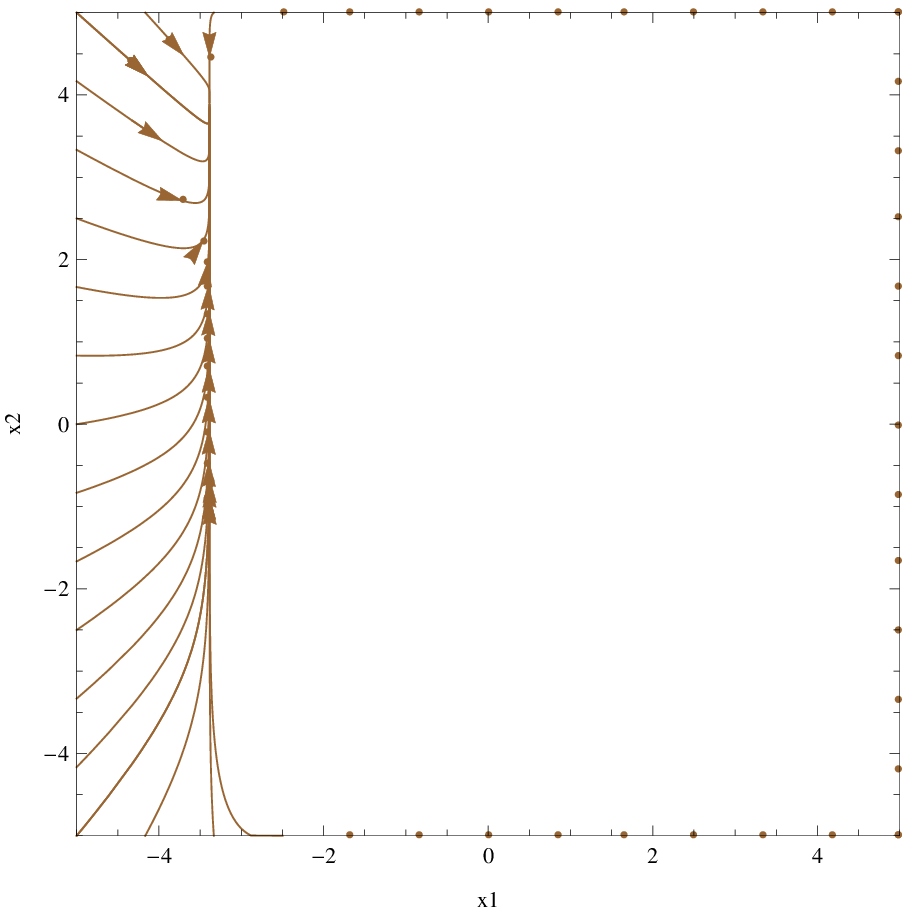}
\caption{{\it{ The vector field flow in the $x_1-x_2$ plane with
$x_3=\frac{1}{2}$ (left plot), and various trajectories in the
$x_1-x_2$ plane, with $m=-9/2$.}}} \label{plot6}
\end{figure}

Let us now demonstrate that the reduced dynamical system  by fixing
$x_3=\frac{1}{2}$, is indeed stable. In Fig. \ref{plot6} we plot the
vector field flow in the $x_1-x_2$ plane by fixing $x_3=\frac{1}{2}$
(left plot), and various trajectories  in the $x_1-x_2$ plane. As it
can be seen in both plots, the fixed point $\phi_*^1$ is reached
asymptotically.
Finally, the strong $x_3$ dependence of the three dimensional system
can be seen in the left and right plots of Fig. \ref{plot7}. As it
can be seen, depending on the initial conditions, the stable fixed
point $\phi_*^1$ may or may not be reached.
\begin{figure}[h]
\centering
\includegraphics[width=20pc]{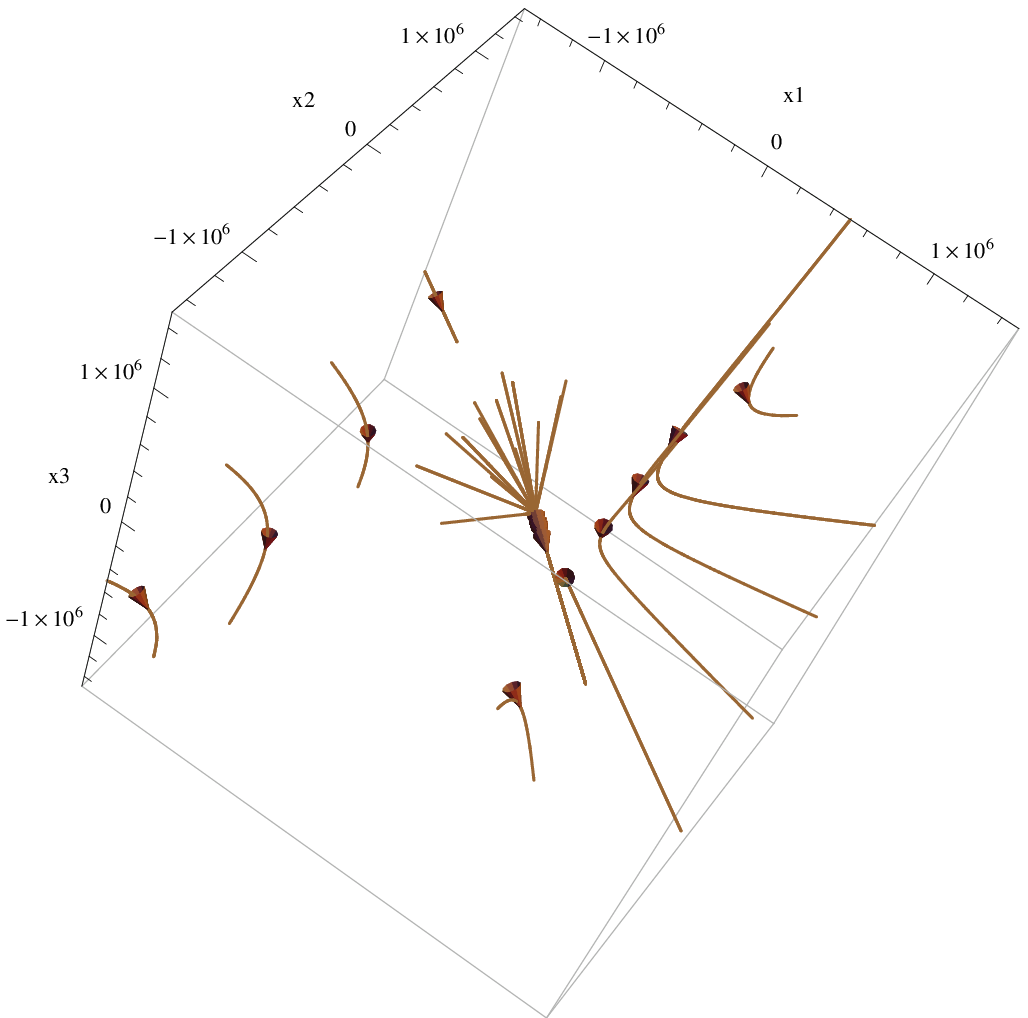}
\includegraphics[width=20pc]{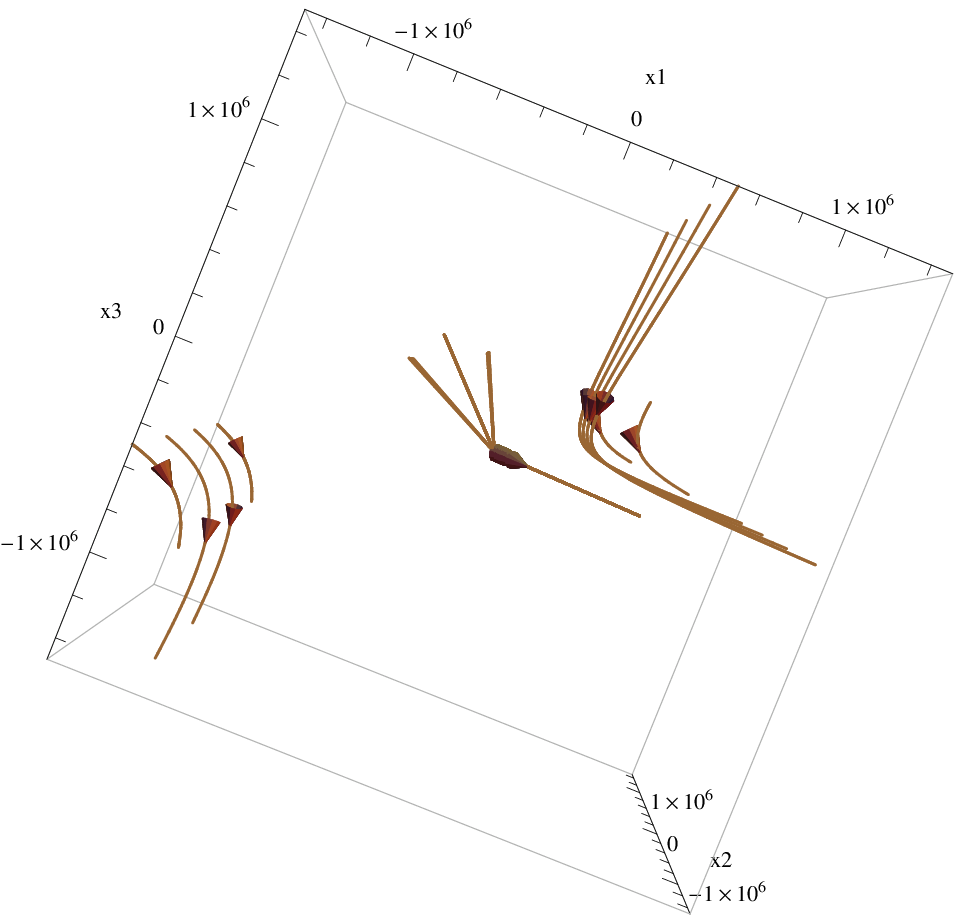}
\caption{{\it{The three dimensional flow with $m=-9/2$.}}}
\label{plot7}
\end{figure}
Having discussed the vacuum $f(R)$ case, what now remains for
completing the study is to add matter and radiation perfect fluids,
and this is the subject of the next section.

Before closing, we need to discuss another interesting point, having to do with the existence of two more fixed points of the dynamical system (\ref{dynamicalsystemmain}) for $m=-9/2$, which are,
\begin{equation}\label{newfixedpoints}
(x_1,x_2,x_3)=(\frac{1}{4} \left(1-i \sqrt{47}\right),\frac{1}{4} \left(-11+i \sqrt{47}\right),\frac{7}{2}),\,\,\,(x_1,x_2,x_3)=(\frac{1}{4} \left(1+i \sqrt{47}\right),\frac{1}{4} \left(-11-i \sqrt{47}\right),\frac{7}{2})\, .
\end{equation}
The form of the new fixed points renders them physically unappealing, since the variables $x_1$ and $x_2$ take complex values. Actually, this means that the $f(R)$ gravity is complex, and this is connected with phantom behavior, as was shown in Ref. \cite{phantom}. Actually, the complex $f(R)$ gravity is usually connected with a phantom scalar theory which develops a Big Rip singularity, and particularly, the complex $f(R)$ gravity corresponds to the region, where the phantom scalar theory develops the crushing type Big Rip singularity. This cannot be accidental, since, as we already mentioned below Eq. (\ref{analyticformx1andx211}) for $x_3=\frac{7}{2}$, the EoS becomes $w_{eff}=-2$, which clearly describes a phantom evolution. We shall not extend the analysis further, since this theory can be useful for a dark energy description, however from an inflationary point of view, it is not so appealing having a phantom theory at hand.

\section{The non-vacuum $f(R)$ Gravity Autonomous Dynamical System: The de Sitter Attractor Case}

In the previous sections we discussed the vacuum $f(R)$ gravity
case, and in this section we shall incorporate the radiation and
matter domination perfect fluids for completeness. We shall focus on
the case $m\simeq 0$, since the other cases can be easily studied.
So for the case $m\simeq 0$ let us analyze the behavior of the phase
space.

Let us first construct the dynamical system in the case that matter
and radiation perfect fluids are considered. The equations of motion
in this case become,
\begin{align}
\label{JGRG15new} 0 =& -\frac{f(R)}{2} + 3\left(H^2 + \dot H\right)
F(R) - 18 \left( 4H^2 \dot H + H \ddot H\right) F'(R)+\kappa^2\rho_{matter}\, ,\\
\label{Cr4b} 0 =& \frac{f(R)}{2} - \left(\dot H + 3H^2\right)F(R) +
6 \left( 8H^2 \dot H + 4 {\dot H}^2 + 6 H \ddot H + \dddot H\right)
F'(R) + 36\left( 4H\dot H + \ddot H\right)^2 F'(R)+
\kappa^2p_{matter}\, ,
\end{align}
where $\rho_{matter}$ and $p_{matter}$ are the total effective
energy density and the effective pressure of the radiation and
matter perfect fluids. In order to reveal the autonomous structure
of the $f(R)$ gravity cosmological system, we introduce the
following variables,
\begin{equation}\label{variablesslowdownnew}
x_1=-\frac{\dot{F}(R)}{F(R)H},\,\,\,x_2=-\frac{f(R)}{6F(R)H^2},\,\,\,x_3=
\frac{R}{6H^2},\,\,\,x_4=\frac{\kappa^2\rho_r}{3FH^2},\,\,\,x_5=\frac{\kappa^2\rho_m}{3FH^2}\,
,
\end{equation}
where $\rho_r$ and $\rho_m$ are the radiation and matter energy
densities. In terms of the $e$-foldings number $N$, the dynamical
system in the case at hand reads,
\begin{align}\label{dynamicalsystemmain2}
&
\frac{\mathrm{d}x_1}{\mathrm{d}N}=-4-3x_1+2x_3-x_1x_3+x_1^2+3x_5+4x_4\,
,
\\ \notag &
\frac{\mathrm{d}x_2}{\mathrm{d}N}=8+m-4x_3+x_2x_1-2x_2x_3+4x_2 \, ,\\
\notag & \frac{\mathrm{d}x_3}{\mathrm{d}N}=-8-m+8x_3-2x_3^2 \, ,\\
\notag & \frac{\mathrm{d}x_4}{\mathrm{d}N}=x_4x_1-2x_4x_3 \, ,,\\
\notag & \frac{\mathrm{d}x_5}{\mathrm{d}N}=x_5+x_5x_1-2x_5x_3 \, ,
\end{align}
where the parameter $m$ is again as it appears in Eq.
(\ref{parameterm}). The matrix
$\mathcal{J}=\sum_i\sum_j\Big{[}\frac{\mathrm{\partial
f_i}}{\partial x_j}\Big{]}$ in the case at hand is equal to,
\begin{equation}\label{matrixceas2}
\mathcal{J}=\left(
\begin{array}{ccccc}
 2 x_1-x_3+3 & 0 & 2-x_1 & 4 & 3 \\
 x_2 & x_1-2 x_3+4 & -2 x_2-4 & 0 & 0 \\
 0 & 0 & 8-4 x_3 & 0 & 0 \\
 x_4 & 0 & -2 x_4 & x_1-2 x_3 & 0 \\
 x_5 & 0 & -2 x_5 & 0 & x_1-2 x_3+1 \\
\end{array}
\right)\, ,
\end{equation}
and in this case the functions $f_i$ for $m\simeq 0$ are,
\begin{align}\label{fis}
& f_1=-4-3x_1+2x_3-x_1x_3+x_1^2+3x_5+4x_4\, , \\ \notag &
f_2=8-4x_3+x_2x_1-2x_2x_3+4x_2\, ,\\ \notag & f_3=8x_3-2x_3^2-8\, ,
\\ \notag & f_4=x_4x_1-2x_4x_3 \, ,\\ \notag & f_5=x_5+x_5x_1-2x_5x_3
\, .
\end{align}
The fixed points of the dynamical system
(\ref{dynamicalsystemmain2}), for $m\simeq 0$ are the following,
\begin{align}\label{fixedpointsgeneral2}
& \phi_*^1=(-1,0,2,0,0),
 \\ \notag & \phi_*^2=(3,0,2,0,-4),
 \\ \notag & \phi_*^3=(4,0,2,-5,0)\, .
\end{align}
The corresponding eigenvalues for the fixed point $\phi_*^1$ are
$(-5, -4, -1, -1, 0)$, while for the fixed point $\phi_*^2$ these
are $(4, 3, 3, -1, 0)$ and finally for the fixed point $\phi_*^3$
the eigenvalues are $(5, 4, 4, 1, 0)$. All the equilibria are
non-hyperbolic and as we now show, only $\phi_*^1$ is stable, which
means that regardless the initial conditions used, the trajectories
are attracted to the fixed point $\phi_*^1$. The fact that $x_3$ is
equal to $x_3=2$ for all the equilibria, shows that these are de
Sitter equilibria, since $w_{eff}=-1$, as in the cases we studied in
the previous section. Let us now focus on the dynamics of the
configuration space spanned by the variables $x_i$, $i=1,...,5$.

The analytical study in this case turns out to be quite complicated, and only the solution for $x_3(N)$ can be found analytically, and this is given in Eq. (\ref{solutionanalyticx3}).

\begin{figure}[h]
\centering
\includegraphics[width=20pc]{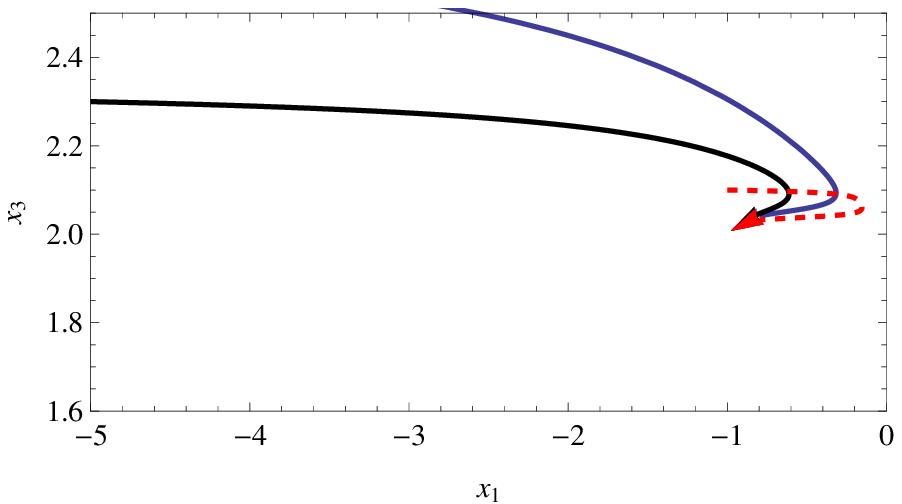}
\includegraphics[width=20pc]{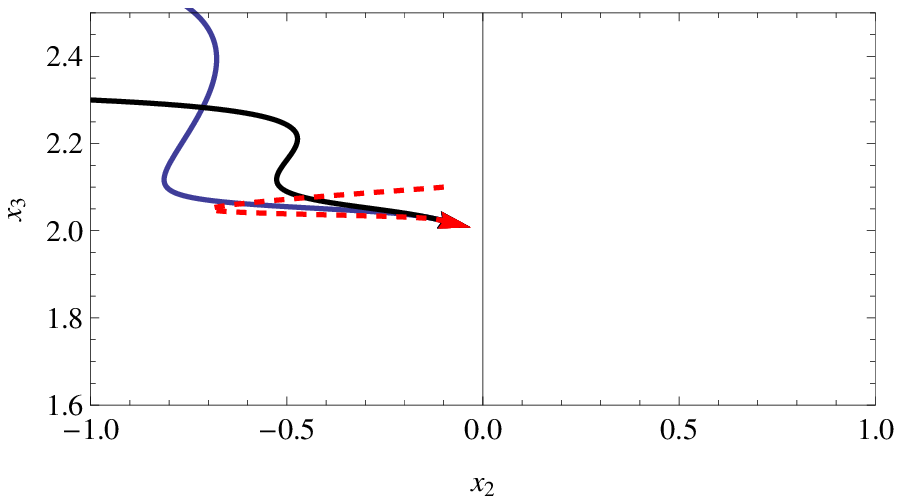}
\includegraphics[width=20pc]{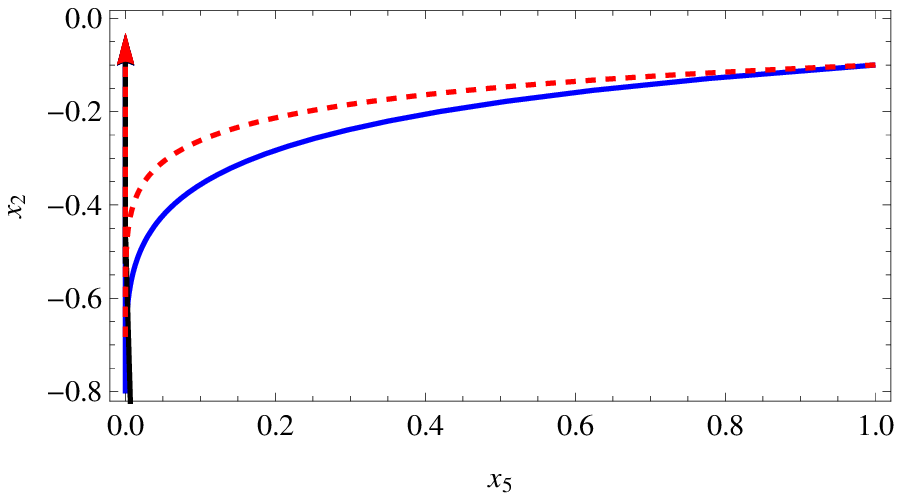}
\caption{{\it{Plots the trajectories of the dynamical system in the
$x_1-x_3$ plane (left plot), in the $x_3-x_2$ plane (right plot),
and in the $x_2-x_5$ plane (bottom plot), for various initial
conditions, for $m\simeq 0$ and for the dynamical system
(\ref{dynamicalsystemmain2})}}} \label{plot8}
\end{figure}

\begin{figure}[h]
\centering
\includegraphics[width=20pc]{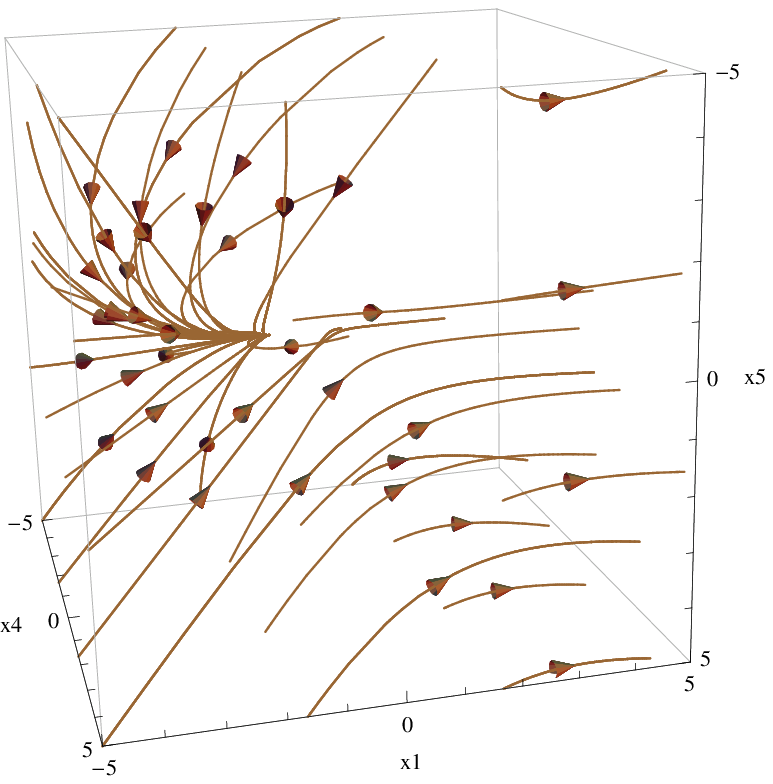}
\includegraphics[width=20pc]{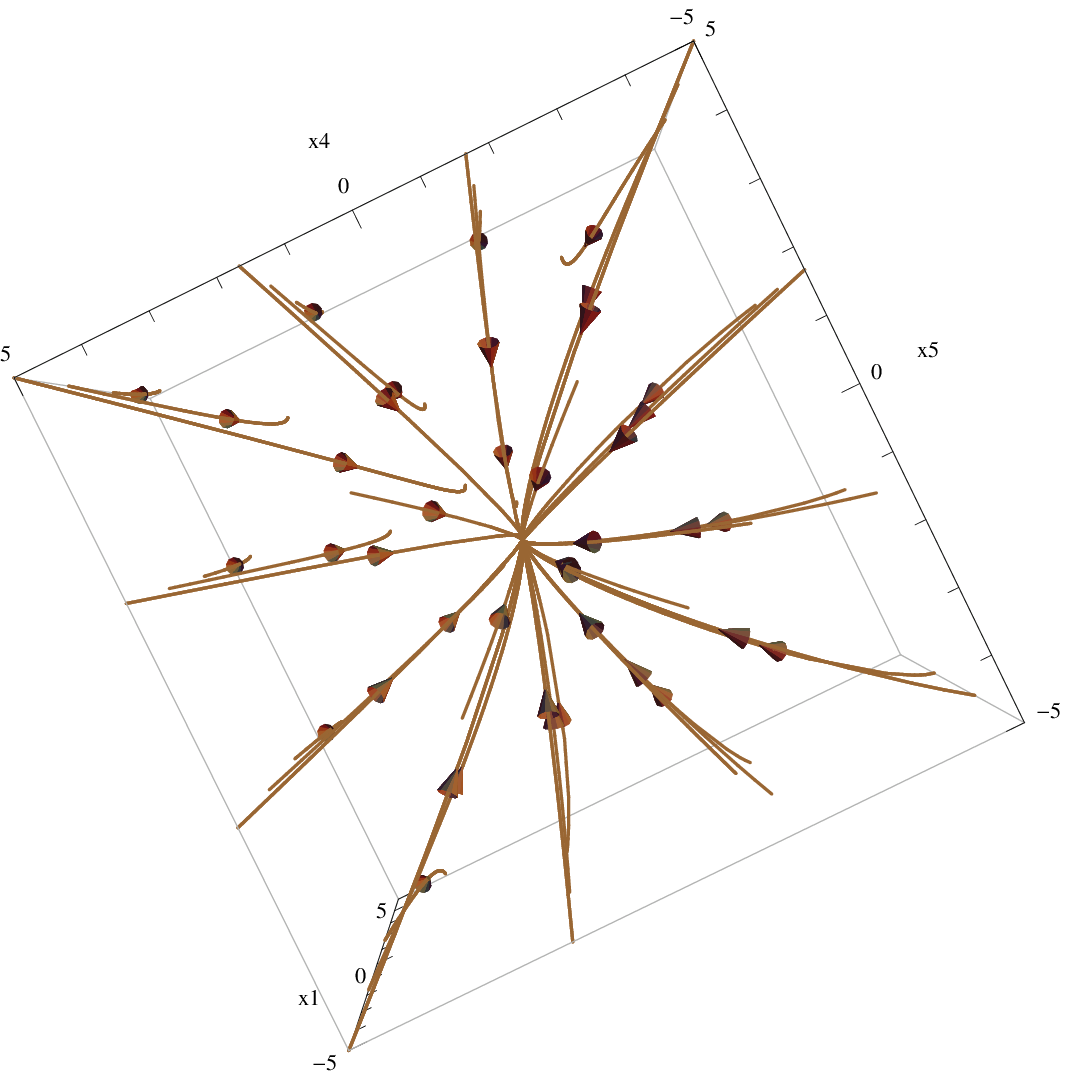}
\caption{{\it{Three dimensional plots of the trajectories in the
space spanned by the variables $(x_1,x_4,_x5)$, by fixing
$(x_2,x_3)$ to their values at the fixed point $\phi_*^1$, that is,
for $(x_2,x_3)=(0,-2)$.}}} \label{plot9}
\end{figure}

Thus we proceed numerically, and after numerically solving the dynamical system
(\ref{dynamicalsystemmain2}) for various sets initial conditions in
Fig. (\ref{plot1})we present the plots of the variables $x_i(N)$
$i=1,..5$, for $N=(0,60)$, by using the initial conditions
$x_1(0)=-10$, $x_2(0)=5$, $x_3(0)=2.6$, $x_4(0)=0.1$ and $x_5(0)=1$.
\begin{figure}[h]
\centering
\includegraphics[width=20pc]{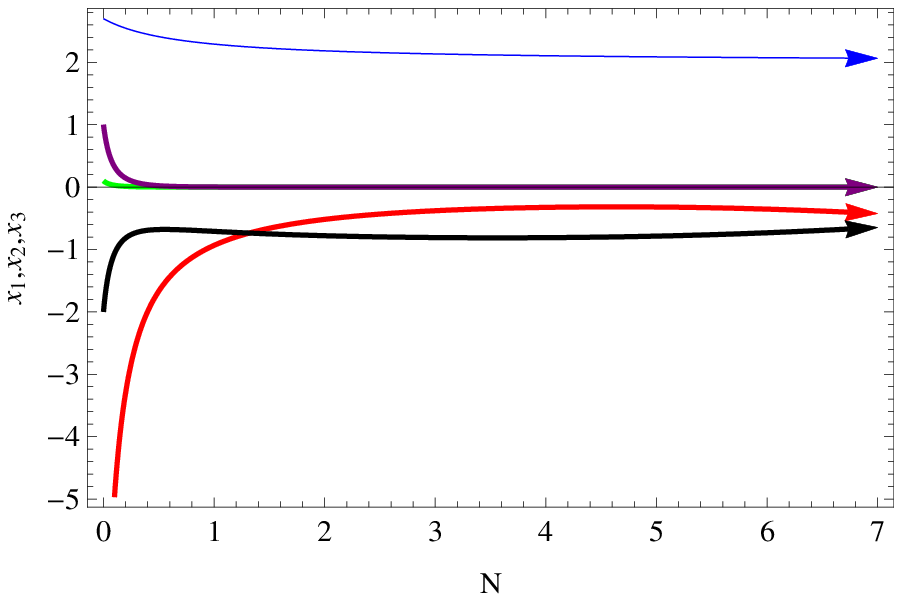}
\includegraphics[width=20pc]{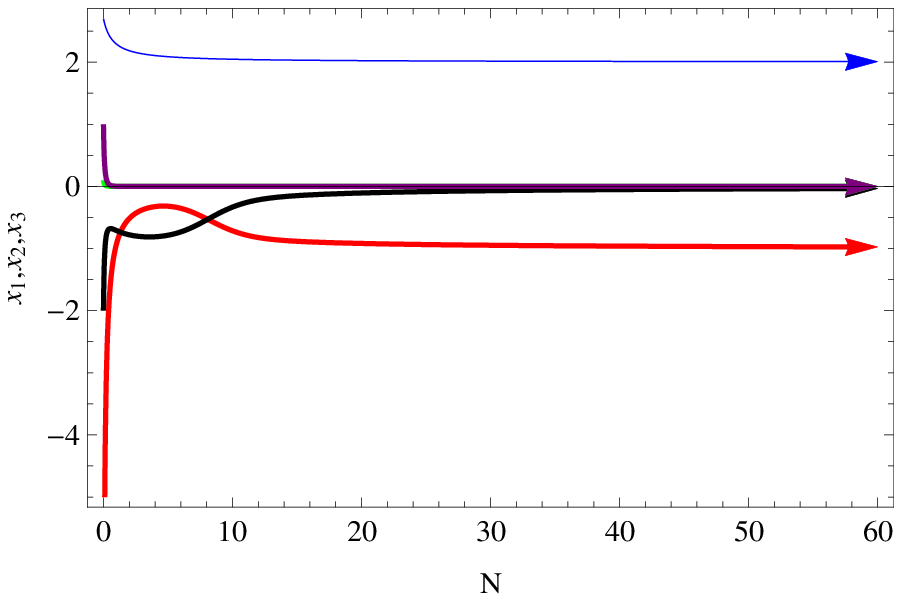}
\caption{{\it{Behavior of the variables $x_i(N)$ $i=1,..5$, for
$N=(0,60)$, by using the initial conditions $x_1(0)=-10$,
$x_2(0)=5$, $x_3(0)=2.6$, $x_4(0)=0.1$ and $x_5(0)=1$, for $m\simeq
0$ and for the dynamical system (\ref{dynamicalsystemmain2})}}}
\label{plot7}
\end{figure}
In Fig. \ref{plot7}, in both plots the red curve is the plot
$x_1-N$, the black curve is $x_2-N$, the blue is $x_3-N$, the green
is $x_4-N$ and finally the purple is $x_5-N$. In the left plot, $N$
belongs in the interval $N=(0,7)$, so the equilibrium $\phi_*^1$ has
not be reached, however for large $N$, that is for $N\sim 50-60$,
the equilibrium fixed point $\phi_*^1$ is finally reached. Hence,
this partially proves the stability, since the same behavior occurs
regardless the initial conditions we use. In order to further
support the stability of the fixed point $\phi_*^1$, in Fig.
(\ref{plot8}), we plot the trajectories of the dynamical system in
the $x_1-x_3$ plane (left plot), in the $x_3-x_2$ plane (right
plot), and in the $x_2-x_5$ plane (bottom plot), for various initial
conditions.
As it can be seen in all the plots of Fig. \ref{plot8}, the
dynamical system approaches asymptotically the fixed point
$\phi_*^1$. Hence it seems that the matter-radiation $f(R)$ gravity
has a global de Sitter attractor. In order to further reveal the
global stability of the attractor $\phi_*^1$, in Fig. \ref{plot9} we
present two 3-dimensional plots of the trajectories in the space
spanned by the variables $(x_1,x_4,x_5)$, by fixing $(x_2,x_3)$ to
their values at the fixed point $\phi_*^1$, that is, for
$(x_2,x_3)=(0,-2)$.
As it can be seen in the left plot but also clearer in the right
plot, viewed by a different angle, the final attractor in the
$(x_1,x_4,x_5)$ configuration space, is the point
$(x_1,x_4,x_5)=(-1,0,0)$.

In conclusion, even when matter and radiation perfect fluids are
added in the $f(R)$ gravity, there is always a global stable de
Sitter attractor of all the cosmologies that satisfy $m\simeq 0$,
exactly as in the vacuum case. This is clearly an inflationary
attractor which is reached asymptotically for $N\sim 50-60$. Also we
need to note that the fixed point is in the point $(x_4,x_5)=(0,0)$,
which is physically correct, since at a de Sitter point, neither the
mass nor the radiation perfect fluids dominate the evolution. In
principle the same study could be repeated for other values of $m$,
like in the previous section, but we refer from going into details
for brevity.

\section{Conclusions}

In this work we performed a detailed analysis of the $f(R)$ gravity
phase space, by using an autonomous dynamical system approach. As we
demonstrated, it is possible to contain all the time-dependence of
the dynamical system in a single variable, which depends only on the
Hubble rate. Hence, for specific values of this parameter, we
investigated the existence and stability of the fixed points. We
mainly focused on the case $m\simeq 0$, and we briefly discussed
also the case $m=-\frac{9}{2}$. For the analysis, we used two forms
of $f(R)$ gravities, firstly the vacuum $f(R)$ gravity and secondly
the $f(R)$ gravity with matter and radiation perfect fluids present.
In the vacuum $f(R)$ gravity case, the case $m\simeq 0$ results to a
stable de Sitter final attractor. We scrutinized the problem of
stability, by using various sets of initial conditions, and the
numerical analysis we performed, supports our claim about the
existence of an asymptotic de Sitter attractor. In fact the
attractor is inflationary as we showed, and it is reached by the
dynamical system at $N\sim 50-60$. The same behavior occurs for the
$f(R)$ gravity in the presence of matter and radiation perfect
fluids. As we demonstrated the dynamical system is attracted to an
asymptotic stable de Sitter attractor, at which $(x_4,x_5)=(0,0)$.
Hence the de Sitter attractor is due to the $f(R)$ gravity solely.

An interesting outcome of the vacuum $f(R)$ gravity case is the
existence of a stable and of an unstable de Sitter attractors.
Actually, as we demonstrated, the approximated form of the $f(R)$
gravity that leads to the unstable de Sitter fixed points is $\sim
R^2$, while the one that leads to the stable is an exponential plus
a cosmological constant. As we discussed, the $R^2$ term is always
related to growing curvature perturbations, which eventually may
cause the graceful exit from inflation. At the dynamical system
level in principle it is not possible to even see the possibility of
graceful exit, since the $f(R)$ gravity is free from ghost
instabilities and also it does not contain the Ostrogradsky
instability \cite{Woodard:2015zca}. The same happens in a scalar
inflationary theory, the de Sitter final attractor is stable at the
level of dynamical system analysis, however, the graceful exit
occurs when the slow-roll parameters become of the order one. In
effect, the graceful exit is triggered by the breakdown of the
slow-roll expansion, which is a perturbative expansion. In some
sense, the dynamics of the graceful exit from inflation cannot be
seen at the configuration space, so extra higher order parameters
are introduced in order to investigate the dynamical evolution at a
perturbative level. Nevertheless, the existence of an $f(R)$ gravity
which may lead to an unstable de Sitter fixed point, may provide
hints that the graceful exit can occur even in pure $f(R)$ gravity.
This result however, is, by all means, non-conclusive since it
provides only hints but no proof. In modified gravities that contain
ghosts, the instability can be seen in the dynamical system level,
even in the configuration space. We have strong hints to believe
this, and in fact in a mimetic $f(R)$ gravity context, there exist
unstable manifolds folding around a stable de Sitter point. The
presence of the unstable manifolds utterly changes the dynamics of a
stable final attractor. We will report on this issue soon, including
the $R^2$ case we discussed in a previous section. The fact that for
the $R^2$ gravity, the dynamical variables tend to the unstable de
Sitter fixed point, indicates the presence of a globally unstable
manifold in the phase space.

\section*{Acknowledgments}

This work is supported by MINECO (Spain), project FIS2013-44881,
FIS2016-76363-P and by CSIC I-LINK1019 Project (S.D.O).


\begin{thebibliography}{99}



\bibitem{Riess:1998cb}
A.~G.~Riess {\it et al.} [Supernova Search Team],
%``Observational evidence from supernovae for an accelerating universe and a cosmological constant,''
Astron.\ J.\ {\bf 116} (1998) 1009
%%%doi:10.1086/300499
[astro-ph/9805201].
%%CITATION = %doi:10.1086/300499;%%
%9542 citations counted in INSPIRE as of 22 Dec 2016





\bibitem{reviews1}
 S.~Nojiri, S.~D.~Odintsov and V.~K.~Oikonomou,
  %``Modified Gravity Theories on a Nutshell: Inflation, Bounce and Late-time Evolution,''
  Phys.\ Rept.\  {\bf 692} (2017) 1
  doi:10.1016/j.physrep.2017.06.001
  [arXiv:1705.11098 [gr-qc]].
  %%CITATION = doi:10.1016/j.physrep.2017.06.001;%%
  %21 citations counted in INSPIRE as of 20 Aug 2017

\bibitem{reviews2}

S. Nojiri, S.D. Odintsov,
   %``Unified cosmic history in modified gravity: from F(R) theory to
   %Lorentz non-invariant models,''
   Phys.\ Rept.\  {\bf 505}, 59 (2011);
   %[arXiv:1011.0544 [gr-qc]].
   %%CITATION = ARXIV:1011.0544;%%


  \bibitem{reviews3}
S. Nojiri, S.D. Odintsov,
  %``Introduction to modified gravity and gravitational alternative for dark
  %energy,''
  eConf {\bf C0602061}, 06 (2006)
  [Int.\ J.\ Geom.\ Meth.\ Mod.\ Phys.\  {\bf 4}, 115 (2007)].
  %[arXiv:hep-th/0601213];
  %%CITATION = 00436,4,115;%%


   \bibitem{reviews4}
 S. Capozziello, M. De Laurentis,
   %``Extended Theories of Gravity,''
   Phys.\ Rept.\  {\bf 509}, 167 (2011);\\
   %[arXiv:1108.6266 [gr-qc]].
   %%CITATION = ARXIV:1108.6266;%%
 V.~Faraoni and S.~Capozziello,
  %``Beyond Einstein Gravity : A Survey of Gravitational Theories for Cosmology and Astrophysics,''
  Fundam.\ Theor.\ Phys.\  {\bf 170} (2010).
  doi:10.1007/978-94-007-0165-6
  %%CITATION = doi:10.1007/978-94-007-0165-6;%%
  %64 citations counted in INSPIRE as of 16 Sep 2017


%\cite{Nojiri:2003ft}
\bibitem{Nojiri:2003ft}
  S.~Nojiri and S.~D.~Odintsov,
  %``Modified gravity with negative and positive powers of the curvature: Unification of the inflation and of the cosmic acceleration,''
  Phys.\ Rev.\ D {\bf 68} (2003) 123512
  doi:10.1103/PhysRevD.68.123512
  [hep-th/0307288].
  %%CITATION = doi:10.1103/PhysRevD.68.123512;%%
  %1239 citations counted in INSPIRE as of 24 Aug 2017


     %\cite{Nojiri:2006gh}
\bibitem{Nojiri:2006gh}
  S.~Nojiri and S.~D.~Odintsov,
  %``Modified f(R) gravity consistent with realistic cosmology: From matter dominated epoch to dark energy universe,''
  Phys.\ Rev.\ D {\bf 74} (2006) 086005
  doi:10.1103/PhysRevD.74.086005
  [hep-th/0608008].
  %%CITATION = doi:10.1103/PhysRevD.74.086005;%%
  %541 citations counted in INSPIRE as of 08 Aug 2017



  %\cite{Boehmer:2014vea}
\bibitem{Boehmer:2014vea}
  C.~G.~Boehmer and N.~Chan,
  %``Dynamical systems in cosmology,''
  doi:10.1142/9781786341044.0004
  arXiv:1409.5585 [gr-qc].
  %%CITATION = doi:10.1142/9781786341044_0004;%%
  %21 citations counted in INSPIRE as of 24 Aug 2017




  %\cite{Bohmer:2010re}
\bibitem{Bohmer:2010re}
  C.~G.~Boehmer, T.~Harko and S.~V.~Sabau,
  %``Jacobi stability analysis of dynamical systems: Applications in gravitation and cosmology,''
  Adv.\ Theor.\ Math.\ Phys.\  {\bf 16} (2012) no.4,  1145
  doi:10.4310/ATMP.2012.v16.n4.a2
  [arXiv:1010.5464 [math-ph]].
  %%CITATION = doi:10.4310/ATMP.2012.v16.n4.a2;%%
  %9 citations counted in INSPIRE as of 24 Aug 2017





  %\cite{Goheer:2007wu}
\bibitem{Goheer:2007wu}
  N.~Goheer, J.~A.~Leach and P.~K.~S.~Dunsby,
  %``Dynamical systems analysis of anisotropic cosmologies in R**n-gravity,''
  Class.\ Quant.\ Grav.\  {\bf 24} (2007) 5689
  doi:10.1088/0264-9381/24/22/026
  [arXiv:0710.0814 [gr-qc]].
  %%CITATION = doi:10.1088/0264-9381/24/22/026;%%
  %45 citations counted in INSPIRE as of 24 Aug 2017





  %\cite{Leon:2014yua}
\bibitem{Leon:2014yua}
  G.~Leon and E.~N.~Saridakis,
  %``Dynamical behavior in mimetic F(R) gravity,''
  JCAP {\bf 1504} (2015) no.04,  031
  doi:10.1088/1475-7516/2015/04/031
  [arXiv:1501.00488 [gr-qc]].
  %%CITATION = doi:10.1088/1475-7516/2015/04/031;%%
  %38 citations counted in INSPIRE as of 24 Aug 2017





  %\cite{Leon:2010pu}
\bibitem{Leon:2010pu}
  G.~Leon and E.~N.~Saridakis,
  %``Dynamics of the anisotropic Kantowsky-Sachs geometries in $R^n$ gravity,''
  Class.\ Quant.\ Grav.\  {\bf 28} (2011) 065008
  doi:10.1088/0264-9381/28/6/065008
  [arXiv:1007.3956 [gr-qc]].
  %%CITATION = doi:10.1088/0264-9381/28/6/065008;%%
  %40 citations counted in INSPIRE as of 24 Aug 2017


  %\cite{deSouza:2007zpn}
\bibitem{deSouza:2007zpn}
  J.~C.~C.~de Souza and V.~Faraoni,
  %``The Phase space view of f(R) gravity,''
  Class.\ Quant.\ Grav.\  {\bf 24} (2007) 3637
  doi:10.1088/0264-9381/24/14/006
  [arXiv:0706.1223 [gr-qc]].
  %%CITATION = doi:10.1088/0264-9381/24/14/006;%%
  %85 citations counted in INSPIRE as of 24 Aug 2017

%\cite{Giacomini:2017yuk}
\bibitem{Giacomini:2017yuk}
  A.~Giacomini, S.~Jamal, G.~Leon, A.~Paliathanasis and J.~Saavedra,
  %``Dynamical Analysis of an Integrable Cubic Galileon Cosmological Model,''
  Phys.\ Rev.\ D {\bf 95} (2017) no.12,  124060
  doi:10.1103/PhysRevD.95.124060
  [arXiv:1703.05860 [gr-qc]].
  %%CITATION = doi:10.1103/PhysRevD.95.124060;%%


%\cite{Kofinas:2014aka}
\bibitem{Kofinas:2014aka}
  G.~Kofinas, G.~Leon and E.~N.~Saridakis,
  %``Dynamical behavior in $f(T,T_G)$ cosmology,''
  Class.\ Quant.\ Grav.\  {\bf 31} (2014) 175011
  doi:10.1088/0264-9381/31/17/175011
  [arXiv:1404.7100 [gr-qc]].
  %%CITATION = doi:10.1088/0264-9381/31/17/175011;%%
  %45 citations counted in INSPIRE as of 24 Aug 2017


%\cite{Leon:2012mt}
\bibitem{Leon:2012mt}
  G.~Leon and E.~N.~Saridakis,
  %``Dynamical analysis of generalized Galileon cosmology,''
  JCAP {\bf 1303} (2013) 025
  doi:10.1088/1475-7516/2013/03/025
  [arXiv:1211.3088 [astro-ph.CO]].
  %%CITATION = doi:10.1088/1475-7516/2013/03/025;%%
  %52 citations counted in INSPIRE as of 24 Aug 2017



%\cite{Gonzalez:2006cj}
\bibitem{Gonzalez:2006cj}
  T.~Gonzalez, G.~Leon and I.~Quiros,
  %``Dynamics of quintessence models of dark energy with exponential coupling to dark matter,''
  Class.\ Quant.\ Grav.\  {\bf 23} (2006) 3165
  doi:10.1088/0264-9381/23/9/025
  [astro-ph/0702227].
  %%CITATION = doi:10.1088/0264-9381/23/9/025;%%
  %37 citations counted in INSPIRE as of 24 Aug 2017



  %\cite{Alho:2016gzi}
\bibitem{Alho:2016gzi}
  A.~Alho, S.~Carloni and C.~Uggla,
  %``On dynamical systems approaches and methods in $f(R)$ cosmology,''
  JCAP {\bf 1608} (2016) no.08,  064
  doi:10.1088/1475-7516/2016/08/064
  [arXiv:1607.05715 [gr-qc]].
  %%CITATION = doi:10.1088/1475-7516/2016/08/064;%%
  %7 citations counted in INSPIRE as of 24 Aug 2017


  %\cite{Biswas:2015cva}
\bibitem{Biswas:2015cva}
  S.~K.~Biswas and S.~Chakraborty,
  %``Interacting Dark Energy in $f(T)$ cosmology : A Dynamical System analysis,''
  Int.\ J.\ Mod.\ Phys.\ D {\bf 24} (2015) no.07,  1550046
  doi:10.1142/S0218271815500467
  [arXiv:1504.02431 [gr-qc]].
  %%CITATION = doi:10.1142/S0218271815500467;%%
  %10 citations counted in INSPIRE as of 24 Aug 2017


  %\cite{Muller:2014qja}
\bibitem{Muller:2014qja}
  D.~Müller, V.~C.~de Andrade, C.~Maia, M.~J.~Rebouças and A.~F.~F.~Teixeira,
  %``Future dynamics in f(R) theories,''
  Eur.\ Phys.\ J.\ C {\bf 75} (2015) no.1,  13
  doi:10.1140/epjc/s10052-014-3227-2
  [arXiv:1405.0768 [astro-ph.CO]].
  %%CITATION = doi:10.1140/epjc/s10052-014-3227-2;%%





%\cite{Mirza:2014nfa}
\bibitem{Mirza:2014nfa}
  B.~Mirza and F.~Oboudiat,
  %``A Dynamical System Analysis of $f(R,T)$ Gravity,''
  Int.\ J.\ Geom.\ Meth.\ Mod.\ Phys.\  {\bf 13} (2016) no.09,  1650108
  doi:10.1142/S0219887816501085
  [arXiv:1412.6640 [gr-qc]].
  %%CITATION = doi:10.1142/S0219887816501085;%%
  %2 citations counted in INSPIRE as of 24 Aug 2017


  %\cite{Rippl:1995bg}
\bibitem{Rippl:1995bg}
  S.~Rippl, H.~van Elst, R.~K.~Tavakol and D.~Taylor,
  %``Kinematics and dynamics of f(R) theories of gravity,''
  Gen.\ Rel.\ Grav.\  {\bf 28} (1996) 193
  doi:10.1007/BF02105423
  [gr-qc/9511010].
  %%CITATION = doi:10.1007/BF02105423;%%
  %25 citations counted in INSPIRE as of 24 Aug 2017


%\cite{Ivanov:2011vy}
\bibitem{Ivanov:2011vy}
  M.~M.~Ivanov and A.~V.~Toporensky,
  %``Cosmological dynamics of fourth order gravity with a Gauss-Bonnet term,''
  Grav.\ Cosmol.\  {\bf 18} (2012) 43
  doi:10.1134/S0202289312010100
  [arXiv:1106.5179 [gr-qc]].
  %%CITATION = doi:10.1134/S0202289312010100;%%
  %8 citations counted in INSPIRE as of 24 Aug 2017


%\cite{Khurshudyan:2016qox}
\bibitem{Khurshudyan:2016qox}
  M.~Khurshudyan,
  %``Phase space analysis in a model of f(T) gravity with nonlinear sign changeable interactions,''
  Int.\ J.\ Geom.\ Meth.\ Mod.\ Phys.\  {\bf 14} (2016) no.03,  1750041.
  doi:10.1142/S0219887817500414
  %%CITATION = doi:10.1142/S0219887817500414;%%
  %1 citations counted in INSPIRE as of 24 Aug 2017


  %\cite{Boko:2016mwr}
\bibitem{Boko:2016mwr}
  R.~D.~Boko, M.~J.~S.~Houndjo and J.~Tossa,
  %``Stability and phase space analysis in $f(R)$ theory with generalized exponential $f(R)$ model,''
  Int.\ J.\ Mod.\ Phys.\ D {\bf 25} (2016) no.10,  1650098
  doi:10.1142/S021827181650098X
  [arXiv:1605.03404 [gr-qc]].
  %%CITATION = doi:10.1142/S021827181650098X;%%


%\cite{Odintsov:2017icc}
\bibitem{Odintsov:2017icc}
  S.~D.~Odintsov, V.~K.~Oikonomou and P.~V.~Tretyakov,
  %``Phase space analysis of the accelerating multifluid Universe,''
  Phys.\ Rev.\ D {\bf 96} (2017) no.4,  044022
  doi:10.1103/PhysRevD.96.044022
  [arXiv:1707.08661 [gr-qc]].
  %%CITATION = doi:10.1103/PhysRevD.96.044022;%%



  %\cite{Odintsov:2015wwp}
\bibitem{Odintsov:2015wwp}
  S.~D.~Odintsov and V.~K.~Oikonomou,
  %``Accelerating cosmologies and the phase structure of F(R) gravity with Lagrange multiplier constraints: A mimetic approach,''
  Phys.\ Rev.\ D {\bf 93} (2016) no.2,  023517
  doi:10.1103/PhysRevD.93.023517
  [arXiv:1511.04559 [gr-qc]].
  %%CITATION = doi:10.1103/PhysRevD.93.023517;%%
  %23 citations counted in INSPIRE as of 24 Aug 2017


\bibitem{dynsystemsbook} Stephen Wiggins, Introduction to Applied Nonlinear Dynamical Systems and
Chaos, Springer, New York, 2003

%\cite{Odintsov:2015ynk}
\bibitem{Odintsov:2015ynk}
  S.~D.~Odintsov and V.~K.~Oikonomou,
  %``Big-Bounce with Finite-time Singularity: The $F(R)$ Gravity Description,''
  Int.\ J.\ Mod.\ Phys.\ D {\bf 26} (2017) no.08,  1750085
  doi:10.1142/S0218271817500857
  [arXiv:1512.04787 [gr-qc]].
  %%CITATION = doi:10.1142/S0218271817500857;%%
  %18 citations counted in INSPIRE as of 24 Aug 2017



%\cite{Elizalde:2010ts}
\bibitem{Elizalde:2010ts}
  E.~Elizalde, S.~Nojiri, S.~D.~Odintsov, L.~Sebastiani and S.~Zerbini,
  %``Non-singular exponential gravity: a simple theory for early- and late-time accelerated expansion,''
  Phys.\ Rev.\ D {\bf 83} (2011) 086006
  doi:10.1103/PhysRevD.83.086006
  [arXiv:1012.2280 [hep-th]].
  %%CITATION = doi:10.1103/PhysRevD.83.086006;%%
  %104 citations counted in INSPIRE as of 15 Sep 2017

%\cite{Bamba:2014jia}
\bibitem{Bamba:2014jia}
  K.~Bamba, R.~Myrzakulov, S.~D.~Odintsov and L.~Sebastiani,
  %``Trace-anomaly driven inflation in modified gravity and the BICEP2 result,''
  Phys.\ Rev.\ D {\bf 90} (2014) no.4,  043505
  doi:10.1103/PhysRevD.90.043505
  [arXiv:1403.6649 [hep-th]].
  %%CITATION = doi:10.1103/PhysRevD.90.043505;%%
  %53 citations counted in INSPIRE as of 15 Sep 2017


\bibitem{starobinsky} A.~A.~Starobinsky,  Phys.\ Lett.\ B {\bf 91} (1980)
99.;\\
J.~D.~Barrow and S.~Cotsakis,  Phys.\ Lett.\ B  214 (1988) 515;
\\J.~D.~Barrow and A.~C.~Ottewill,
  %``The Stability of General Relativistic Cosmological Theory,''
  J.\ Phys.\ A {\bf 16} (1983) 2757.
  doi:10.1088/0305-4470/16/12/022
  %%CITATION = doi:10.1088/0305-4470/16/12/022;%%
  %312 citations counted in INSPIRE as of 16 Sep 2017

%\cite{Odintsov:2015gba}
\bibitem{Odintsov:2015gba}
  S.~D.~Odintsov and V.~K.~Oikonomou,
  %``Singular Inflationary Universe from $F(R)$ Gravity,''
  Phys.\ Rev.\ D {\bf 92} (2015) no.12,  124024
  doi:10.1103/PhysRevD.92.124024
  [arXiv:1510.04333 [gr-qc]].
  %%CITATION = doi:10.1103/PhysRevD.92.124024;%%
  %31 citations counted in INSPIRE as of 14 Sep 2017


\bibitem{phantom}

  F.~Briscese, E.~Elizalde, S.~Nojiri and S.~D.~Odintsov,
  %``Phantom scalar dark energy as modified gravity: Understanding the origin of the Big Rip singularity,''
  Phys.\ Lett.\ B {\bf 646} (2007) 105
  doi:10.1016/j.physletb.2007.01.013
  [hep-th/0612220].
  %%CITATION = doi:10.1016/j.physletb.2007.01.013;%%
  %189 citations counted in INSPIRE as of 27 Oct 2017


%\cite{Woodard:2015zca}
\bibitem{Woodard:2015zca}
  R.~P.~Woodard,
  %``Ostrogradsky's theorem on Hamiltonian instability,''
  Scholarpedia {\bf 10} (2015) no.8,  32243
  doi:10.4249/scholarpedia.32243
  [arXiv:1506.02210 [hep-th]].
  %%CITATION = doi:10.4249/scholarpedia.32243;%%
  %99 citations counted in INSPIRE as of 16 Sep 2017
\end{thebibliography}
\end{document}